\newcommand{\beq}{\begin{equation}}
\newcommand{\eeq}{\end{equation}}

\newcommand{\bear}{\begin{eqnarray}}
\newcommand{\eear}{\end{eqnarray}}

\newcommand{\tn}{\textnormal}

\documentclass[review]{elsart5p}

\usepackage{graphicx}
\usepackage{enumerate}
\usepackage{pifont}
\usepackage{esvect}
\usepackage{amssymb}
\usepackage{bbding}
\usepackage{lineno}
\usepackage{amsmath}
\usepackage{color}
\usepackage[dvipsnames]{xcolor}
\usepackage{dsfont}
\definecolor{col1}{rgb}{0.858, 0.188, 0.478}
\begin{document}

\begin{frontmatter}

\title{Microscopic simulation of xenon-based optical TPCs in the presence of molecular additives}

\author[1] {C.D.R. Azevedo},
\author[3] {D. Gonz\'alez-D\'iaz\corauthref{aut1}},
\corauth[aut1]{Corresponding author.}\ead{Diego.Gonzalez.Diaz@usc.es}
\author[4] {S. F. Biagi},
\author[5] {C.A.B. Oliveira},
\author[6] {C.A.O. Henriques},
\author[7] {J. Escada},
\author[22] {F. Monrabal},
\author[8] {J.J. G\'omez-Cadenas\corauthref{aut2}},
\corauth[aut2]{Spokesperson.}
\author[8] {V. \'Alvarez},
\author[8] {J. M. Benlloch-Rodr\'iguez}
\author[7] {F.I.G.M. Borges},
\author[8] {A. Botas},
\author[8] {S. C\'arcel},
\author[8] {J. V. Carri\'on},
\author[10] {S. Cebri\'an},
\author[7] {C.A.N. Conde},
\author[8] {J. D\'iaz},
\author[21] {M. Diesburg},
\author[12] {R. Esteve},
\author[8] {R. Felkai},
\author[6] {L.M.P. Fernandes},
\author[8] {P. Ferrario},
\author[13] {A.L. Ferreira},
\author[6] {E.D.C. Freitas},
\author[5] {A. Goldschmidt},
\author[9] {R.M. Guti\'errez},
\author[14] {J. Hauptman},
\author[9] {A. I. Hernandez},
\author[3] {J.A. Hernando Morata},
\author[12] {V. Herrero},
\author[22] {B.J.P. Jones},
\author[16] {L. Labarga},
\author[8] {A. Laing},
\author[21] {P. Lebrun},
\author[8] {I. Liubarsky},
\author[8] {N. Lopez-March},
\author[9] {M. Losada},
\author[8] {J. Mart\'in-Albo}
\textsuperscript{\!\!,1}\thanks{Now at University of Oxford, United Kingdom.},
\author[3] {G. Mart\'inez-Lema},
\author[8] {A. Mart\'inez},
\author[22] {A. D. McDonald},
\author[6]  {C.M.B. Monteiro},
\author[12] {F.J. Mora},
\author[13] {L.M. Moutinho},
\author[8] {J. Mu\~noz Vidal},
\author[8] {M. Musti},
\author[8] {M. Nebot-Guinot},
\author[8] {P. Novella},
\author[22] {D. Nygren}
\textsuperscript{\!\!,2}\thanks{Co-spokesperson.},
\author[8]  {B. Palmeiro},
\author[21] {A. Para},
\author[16] {J. P\'erez},
\author[8] {M. Querol},
\author[8] {J. Renner},
\author[18] {L. Ripoll},
\author[8] {J. Rodr\'iguez},
\author[22] {L. Rogers},
\author[7]{F.P. Santos},
\author[6]{J.M.F. dos Santos},
\author[8] {L. Serra},
\author[5] {D. Shuman},
\author[8] {A. Sim\'on},
\author[19] {C. Sofka}
\textsuperscript{\!\!,3}\thanks{Now at University of Texas at Austin, USA.},
\author[8] {M. Sorel},
\author[19] {T. Stiegler},
\author[12] {J.F. Toledo},
\author[18] {J. Torrent},
\author[20] {Z. Tsamalaidze},
\author[13]{J.F.C.A. Veloso},
\author[19] {R. Webb},
\author[19] {J.T. White}
\textsuperscript{\!\!,4}\thanks{Deceased.},
\author[8] {N. Yahlali},

\address[1]{I3N - Physics Department, University of Aveiro, Aveiro, Portugal}
\address[3]{Instituto Galego de F\'isica de Altas Enerx\'ias, Universidade de Santiago de Compostela, Santiago de Compostela, Spain}
\address[4]{Uludag University, Faculty of Arts and Sciences Physics Department, Bursa, Turkey}
\address[5]{Lawrence Berkeley National Laboratory, California, USA}
\address[22] {Department of Physics, University of Texas at Arlington Arlington, Texas 76019, USA}
\address[6]{LIBPhys, Departamento de F\'isica, Universidade de Coimbra, Coimbra, Portugal}
\address[7]{Departamento de F\'isica, Universidade de Coimbra, Coimbra, Portugal}
\address[8]{Instituto de F\'isica Corpuscular (IFIC), CSIC \& Universitat de Val\`encia, Valencia, Spain}
\address[9]{Centro de Investigaciones en Ciencias B\'asicas y Aplicadas, Universidad Antonio Nari\~no, Bogot\'a, Colombia}
\address[10]{Laboratorio de F\'isica Nuclear y Astropart\'iculas, Universidad de Zaragoza, Zaragoza, Spain}
\address[21]{Fermi National Accelerator Laboratory Batavia, Illinois, USA}
\address[12]{Instituto de Instrumentaci\'on para Imagen Molecular (I3M), CSIC-Universitat Polit\`ecnica de Val\`encia, Valencia, Spain}
\address[13]{Institute of Nanostructures, Nanomodelling and Nanofabrication (i3N), Universidade de Aveiro, Aveiro, Portugal}
\address[14]{Department of Physics and Astronomy, Iowa State University, Iowa, USA}
\address[16]{Departamento de F\'isica Te\'orica, Universidad Aut\'onoma de Madrid, Madrid, Spain}
\address[17]{Dpto.\ de Mec\'anica de Medios Continuos y Teor\'ia de Estructuras, Univ.\ Polit\`ecnica de Val\`encia, Valencia, Spain}
\address[18]{Escola Polit\`ecnica Superior, Universitat de Girona, Girona, Spain}
\address[19]{Department of Physics and Astronomy, Texas A\&M University, Texas, USA}
\address[20]{Joint Institute for Nuclear Research (JINR), Dubna, Russia}

\begin{abstract}
We introduce a simulation framework for the transport of high and low energy electrons in xenon-based optical time projection chambers (OTPCs). The simulation relies on elementary cross sections (electron-atom and electron-molecule) and incorporates, in order to compute the gas scintillation, the reaction/quenching rates (atom-atom and atom-molecule) of the first 41 excited states of xenon and the relevant associated excimers, together with their radiative cascade. The results compare positively with observations made in pure xenon and its mixtures with CO$_2$ and CF$_4$ in a range of pressures from 0.1 to 10~bar. This work sheds some light on the elementary processes responsible for the primary and secondary xenon-scintillation mechanisms in the presence of additives, that are of interest to the OTPC technology.
\end{abstract}

\end{frontmatter}

% main text

\section{Introduction}
\label{intro}

Gas-based detectors belonging to the family of optical time projection chambers (or OTPCs, a term coined in \cite{Dominik1}) are the workhorse of a number of modern experiments, including those devoted to the study of the 2-proton decay \cite{Dominik2}, excited Hoyle states \cite{Zimmerman}, neutrino-less double-beta decay \cite{DEMO1, DEMO2} or `directional' dark matter detection \cite{DDM1, DDM2}. For small systems, OTPCs can presently outperform classical charge-readout TPCs in simplicity and robustness \cite{Margato,Florian,Fil}, but they can as well, in special configurations, significantly enhance the topological \cite{DDM2} or calorimetric \cite{DEMO1} information of the event under study. Natural scintillators like TMA, TMAE and TEA were the first to be used for imaging particle tracks in gaseous detectors in the late 80's \cite{Char1}, however modern OTPCs rely mostly on the scintillation of just three common additives: N$_2$, CF$_4$, and Xe, effective either as wavelength-shifters or as the main scintillating gas.

This wealth of applications and increasing interest has not led, apparently, to a comparable effort on the theoretical front. %except for the recent computation of secondary scintillation yields in pure noble gases \cite{Oli1, Oli2}.
We concentrate in this work on a particularly relevant case: the scintillation of xenon in the presence of molecular additives in amounts capable of beneficially altering the electron swarm dynamics. We do so by resorting to a fully microscopic description of the process, with the aim of elucidating the underlying mechanisms of gaseous scintillation in conditions of practical interest to detector builders.
%At doing so, we extend as well those early simulations in order to make them capable of computing the wavelength and time characteristics of the emission, by resorting to a fully microscopic picture.

Microscopic simulations are relevant for particle detectors with optical readout provided their ultimate output are the space-time coordinates of one (or several) photons emitted from each excited seed state, together with the associated ionization electrons. A microscopic approach can minimize pitfalls on assumptions about how the particle's energy is distributed throughout the observable scintillation spectrum and at which position and time. For the case of xenon, microscopic simulations have been successfully developed during the 80's for the evaluation of the xenon excimer laser \cite{Lorents, Eckstrom, Thomson}, and they are nowadays ubiquitous for instance in the study of xenon discharges, when coupled to the hydrodynamic equations \cite{Gnybida,DiZhu}. It is customary in those cases to focus on the global system response upon an initial `macroscopic' energy release (e.g, an electron gun shot or a plasma ignition). This leads invariably to scenarios approaching conditions of full charge recombination, where electron interactions with highly excited or ionized states play an important role too, as is characteristic of plasmas. None of these conditions apply to particle detectors under standard operation, for which the response to individual particles conveys the primordial information, and interactions take place largely with ground-state neutral species (except for the thermal agitation). Moreover, as it will be shown, the de-excitation pathways can be a priori followed in greater detail in this latter case, by using specific state-to-state quenching rate constants, radiative transition coefficients and excimer formation rates, that are nowadays known to a large degree for the first s, p and d multiplets of xenon.

\begin{figure}[h!!!]
\centering
\includegraphics*[width=\linewidth]{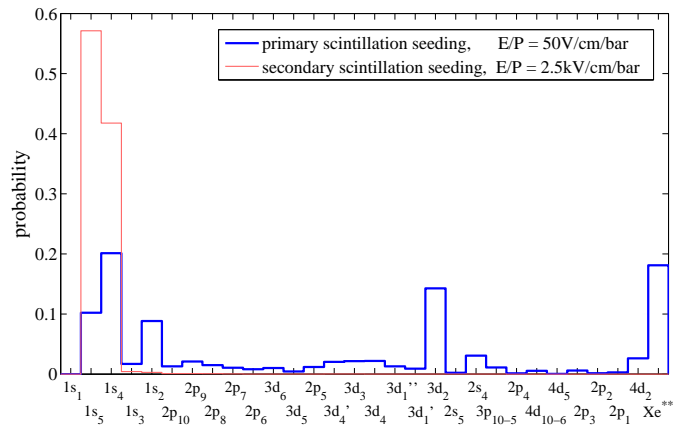}
\caption{Simulated probability distributions for the first 41 excited states of xenon upon interaction of a 30 keV electron. The primary scintillation seeding (thick lines) refers to the case where the excited states are directly promoted by the impinging electron; in the secondary scintillation case the seeding is generated by the ionization electrons, with assistance from an external field at a convenient location (thin lines).}
\label{states}
\end{figure}

Two fundamental seeding processes are of importance to particle detectors: i) the bottom-up seeding characteristic of secondary scintillation (e.g. \cite{Policarpo}), where thermalized ionization electrons gain energy in an external field and, in the case of xenon, they go on to predominantly populate the lowest lying metastable 1s$_5$-6s$[3/2]_2$($^3$P$_2$) and resonant 1s$_4$-6s$[3/2]_1$($^3$P$_1$) atomic excited states;\footnote{Throughout the text, Paschen notation is used for atoms. Racah and spin-orbit coupling notations may be included with a hyphen and in brackets, respectively, for some relevant states. The interested reader can find additional information in table \ref{TableI} and \cite{Notation}.} and ii) the top-down seeding characteristic of the excitation created by an energetic primary particle, that typically displays a fairly democratic distribution of excited states followed by a fast radiative and collisional cascade.\footnote{Other common sources of scintillation like charge recombination and Cherenkov light are comparatively small in the conditions discussed hereafter.} The results for these two types of seeding are illustrated in Fig. \ref{states} for 30 keV primary electrons.

To the authors' knowledge, the first microscopic 3D-simulation of relevance to scintillating gaseous detectors was introduced by Dias and Santos in the 90's for the computation of secondary scintillation in pure xenon and xenon-neon mixtures \cite{Filo0,Filo1, Filo2}, under the simplifying assumption that any excited state would yield a photon in the xenon second continuum. As it will be shown, under typical conditions this represents a valid proxy with an accuracy higher than 5\% and, importantly, it allows neglecting the microphysics of the atomic cascade. An up-to-date and open-source simulation software, including the first 60 excited xenon states, has been recently introduced in \cite{Oli1, Oli2}, extending this early work to other noble gases as well as to the infrared scintillation.

The first attempt to include the presence of molecular additives enforced the introduction of 2 and 3-body molecular quenching rates, and some coarse prescription for the cascade, a pioneering work done by Escada et al. in \cite{Escada}. This spurred subsequent experimental work in \cite{homeo}, during which the limitations of the original simulation approach and the necessity of a more refined treatment started to become apparent.

\begin{figure}[h!!!]
\centering
\includegraphics*[width=\linewidth]{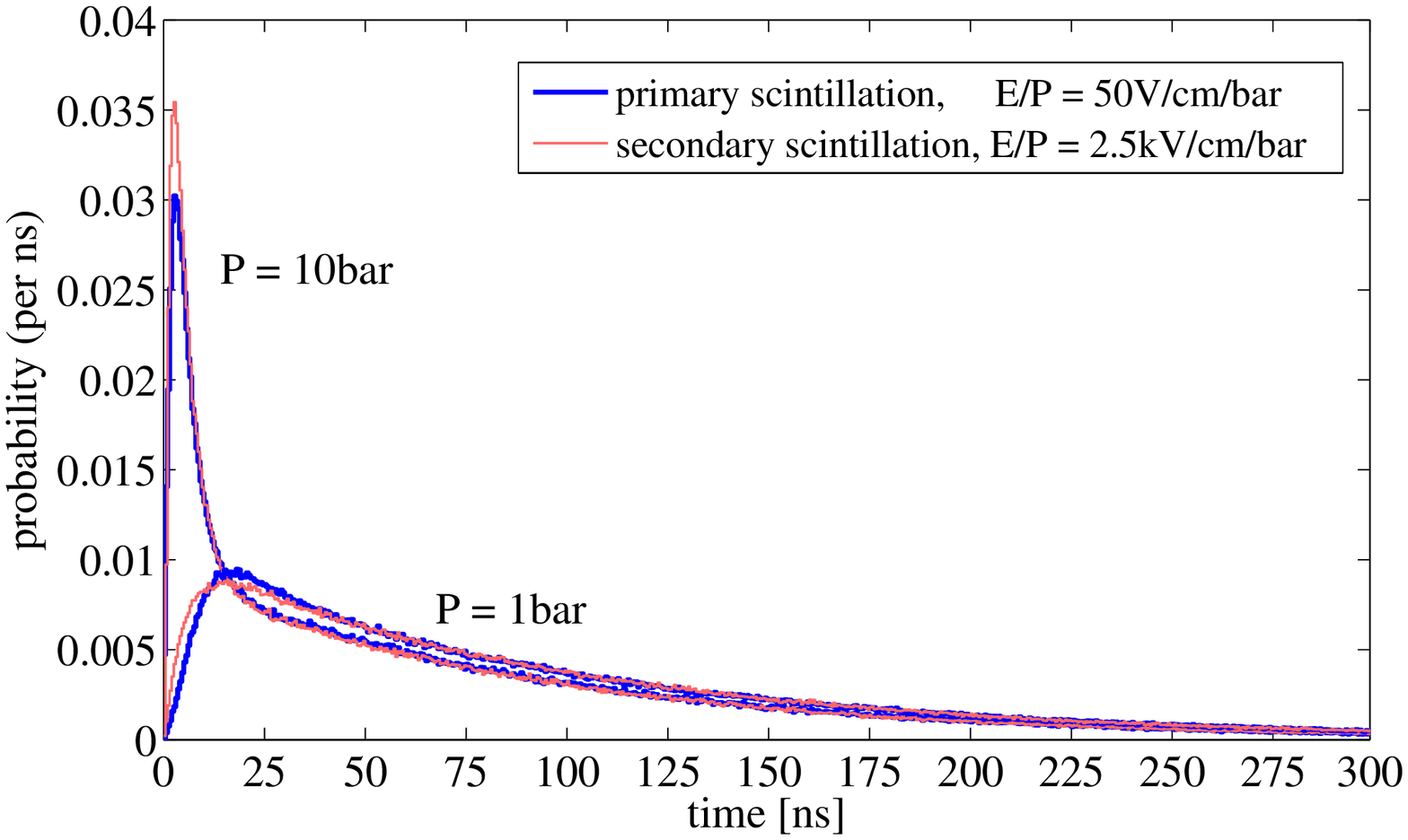}

\includegraphics*[width=\linewidth]{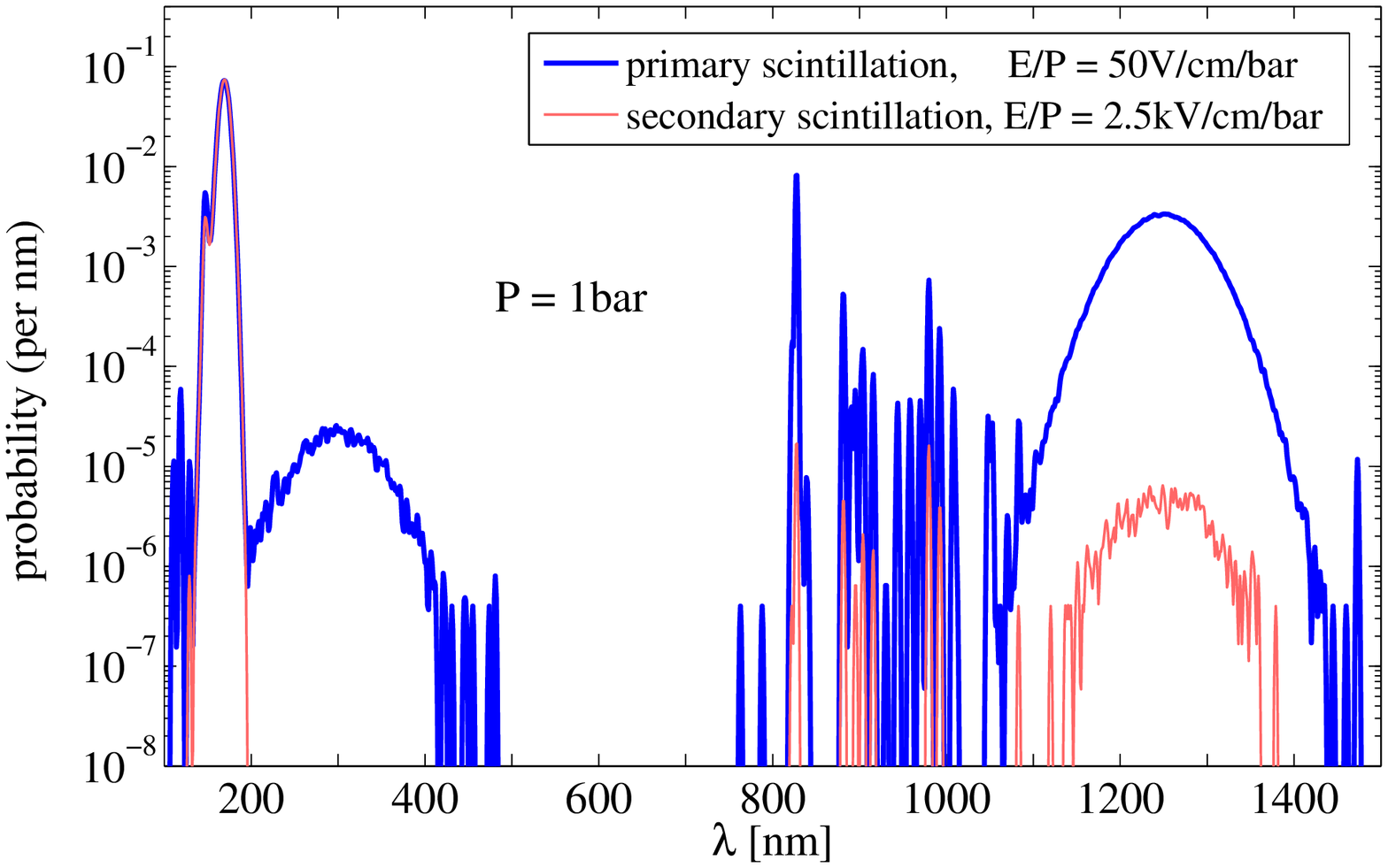}
\caption{Simulated time (top) and wavelength (bottom) distributions for the primary (thick lines) and secondary (thin lines) scintillation, upon interaction of 30 keV electrons at 1 and 10 bar. The upper distribution corresponds to the scintillation for the so-called xenon `second continuum' ($155\tn{nm}< \lambda <195\tn{nm}$).}
\label{EL_30keV_spect}
\end{figure}

It is important to note that phenomenological models popularized for dark matter detectors, like NEST \cite{NEST}, can, according to the authors, compute the scintillation in noble gases too. At the moment NEST does not contain, however, the microphysics necessary to predict the scintillation characteristics in the presence of additives. For illustrative purposes, Fig. \ref{EL_30keV_spect} shows some representative results of the present simulation framework, obtained after the seeding given by Fig. \ref{states}: the primary (thick lines) and secondary (thin lines) scintillation characteristics for 30 keV electrons are given.

The present work has been developed within the NEXT quest for enhanced scintillating mixtures, specifically those that can maintain the xenon VUV-scintillation at usable levels while reducing the electron diffusion from 10 mm to the 2 mm scale. The importance of such an asset for an enhanced topological signature in the reconstruction of a hypothetical $\beta\beta0\nu$ decay has been recently demonstrated in \cite{Josh}. Besides introducing (section \ref{Model}) and benchmarking (section \ref{PureXe}) the model, and comparing with recent data (section \ref{MixXe}), the basic ingredients behind this technological possibility will be discussed on section \ref{Pdep}. A theoretical description of more convoluted approaches akin to the Penning-Fluorescent Xe-TMA TPC described in \cite{PenningFluoro,Accurate,Yasu} falls outside the considerations of the present work, but will be performed in the future.

\section{Methodology} \label{Model}

\subsection{Electron transport}\label{Tranport}

Electron transport simulations presented in this work rely on the codes Degrad \cite{Magboltz} and Garfield++ \cite{Garfield++}, which are interfaced with the Magboltz electron-atom/molecule cross section database \cite{Magboltz}. Garfield++ is optimized for the transport of low energy electrons ($\lesssim 100$ eV) in arbitrary geometries and fields, as is of relevance for engineering gaseous detectors. Degrad makes use of the same set of cross sections extrapolated to high electron energy (MeV scale) by using the Born approximation; currently, it assumes uniform fields and infinite volume. Besides doing the transport, both programs keep track of the population of excited states, that can be used later on as seeds for the computation of the scintillation. The self-field originated by the free charges is not included, therefore any space-charge effect and in particular charge recombination cannot be computed at the moment. According to the measurements performed in \cite{Balan}, for electrons as primaries, this assumption is valid in pure xenon up to 10 bar as long as reduced fields are not much below $E/P = 10$ V/cm/bar.

\begin{figure}[h!!!]
\centering
\includegraphics*[width=\linewidth]{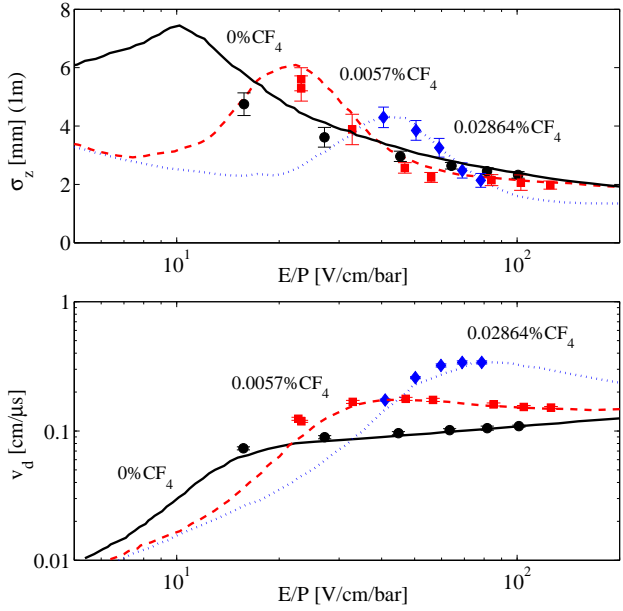}
\caption{Drift velocity (v$_d$) and longitudinal charge spread during 1m-drift ($\sigma_z$) obtained for Xe-CF$_4$ mixtures in the NEXT-DBDM prototype at 10 bar, following the analysis method described in \cite{DBDM}. Lines show simulation results obtained by using the Magboltz cross section database.}
\label{vdDl}
\end{figure}

Recently, the ionization cloud stemming from the transport of energetic electrons and X-rays up to 60 keV has been studied with Degrad through position resolution data for argon and xenon in the range 1-10 bar, showing good agreement \cite{Aze1, Aze2}. The program is benchmarked as well with electron cloud size data from Kobetich and Katz \cite{Kobetich}, in the range 100eV-1MeV. The computed energy to produce an electron-ion pair $W_I$ and the Fano factor $F$ in xenon are $W_I=22.5 \pm 0.2$ eV and $F=0.17$ for electrons above 1 keV, in agreement with present estimates \cite{ICRU, Aprile}. Concerning the transport of low energy `swarm' electrons, the Magboltz database represents the current standard for the electron transport in gaseous detectors. As an example, Fig. \ref{vdDl} provides computed and experimental results obtained by this collaboration for Xe+CF$_4$ mixtures at 10~bar.  Comparisons for the case of Xe-TMA mixtures in the range 1-10bar can be found in \cite{Accurate} and references therein.

\begin{figure}[h!!!]
\centering
\includegraphics*[width=7.8cm]{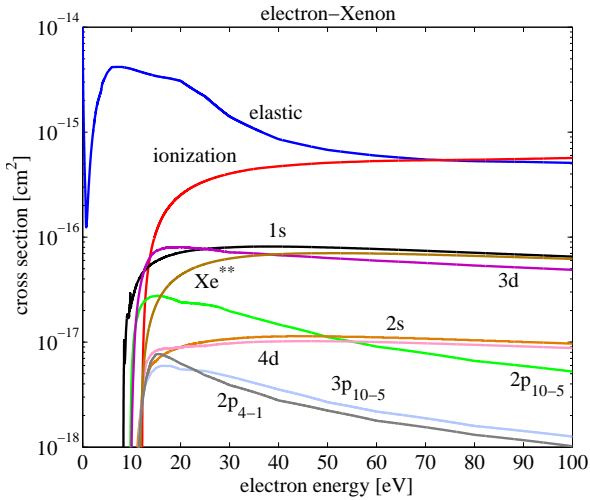}
\caption{Electron-xenon cross sections used in this work. The cross sections for populating the different states have been grouped in multiplets and sub-multiplets for representation. The Magboltz database contains at the moment around 60 xenon excited states, however we restrict this simulation to 41, and include the rest as an effective Xe$^{**}$ state.}
\label{XsectionsXenon}
\end{figure}

\begin{figure}[h!!!]
\centering
\includegraphics*[width=\linewidth]{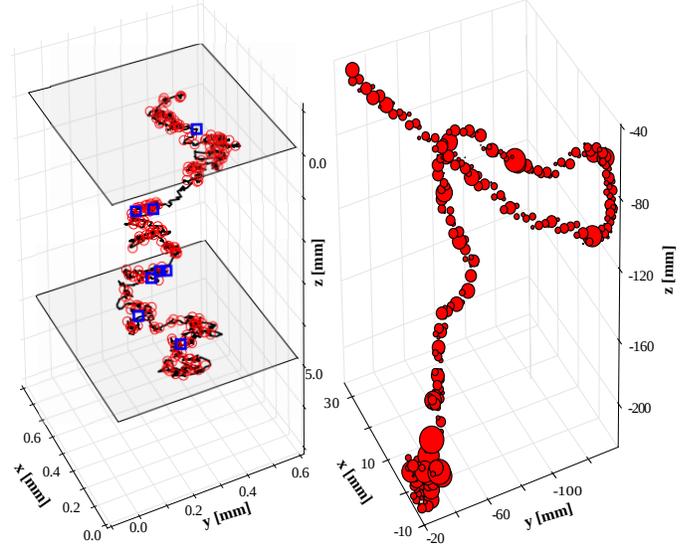}
\caption{Left: trajectory of a single ionization electron (black line) together with the positions of the 1s$_4$ and 1s$_5$ Xe states (circles) and high lying ones (squares), computed with Garfield++. The electron is assumed to be thermalized at the initial position, and it traverses 5 mm gas gap at 10bar and E/P=2.5 kV/cm/bar. Right: high energy electron (2.4 MeV) at E/P=50 V/cm/bar and 10bar, computed with Degrad, with spheres being proportional to the density of excited states at each position. Electrons were released in both cases from the top of the figure.}
\label{TrackFigs}
\end{figure}

The Magboltz database makes use of a subjective evaluation from 1 to 5 in order to indicate how well a given gas is described, yielding for the gases studied here the maximum rating. Electron-xenon cross sections are shown in Fig. \ref{XsectionsXenon}, once grouped in multiplets and sub-multiplets for clarity. Code versions used are 10.14(Garfield++) and 2.14(Degrad), whose outputs are illustratively depicted in Fig. \ref{TrackFigs}.

\subsection{Scintillation}

In order to simulate the scintillation process, a stand-alone MATLAB-based simulation package was developed. It computes the atomic and excimer cascade by resorting to the Einstein coefficients ($A_{ij}$) of xenon and to the 2- and 3-body quenching rates for the processes Xe$^*$+Xe, Xe$^*$+Xe+Xe, Xe$^*$+M and Xe$^*$+Xe+M (where M represents an additive). Higher order reactions like Xe$^*$+M+M can be safely neglected for the sub-\% concentration regime studied here. We conveniently restricted ourselves in this first version of the simulation code to common additives that display a very high transparency to the xenon 2$^{\tn{nd}}$ continuum (CO$_2$, CF$_4$, CH$_4$, CO, H$_2$, N$_2$) plus common impurities like O$_2$ and H$_2$O.

\subsubsection{Atomic cascade} \label{AtomicCascade}

Forty one xenon excited states are included in the simulation, with their corresponding cross sections (Fig. \ref{XsectionsXenon}). The remaining oscillator strength is assigned to an effective Xe$^{**}$ state, whose population can amount to up to 18\% of the excited states responsible for the primary scintillation. Once the populations of all states have been determined during electron transport (section \ref{Tranport}), the radiative and collisional atomic cascade proceeds.

De-excitation through 2-body reactions with ground-state atoms is the dominant cascade mechanism above some 10's of mbar in xenon; the involved rate constants have been measured for instance in \cite{Setser86,Bowering86,Bruce90,Alford92,Whitehead95,Alekseev99}. Information about the end-states resulting from such binary encounters is partly available from those works, in particular for the 1s orbitals, eight of the ten states in the 2p multiplet, three states in 3p$_{10-5}$ and one in 3d. In the absence of state-to-state measurements for the same multiplet, intra-multiplet maximal mixing with detailed balance can be used \cite{Bowering86}. A full inter-multiplet coupling was assumed for the 3d-4d multiplets, by analogy with the 2p-3p case \cite{Whitehead95}. In the few cases where the global information about the quenching rates was missing (2s and Xe$^{**}$ states) values were taken from the average of the 3d and 3p states, and from the 3p states alone, respectively.

Three-body collisions become important for the 1s$_4$, 1s$_5$ \cite{Moutard} and 2p$_5$ \cite{Moutard86} states already above 100 mbar. Due to their role as gateways for the VUV and IR excimer emission, the ensuing pathways are discussed separately in the next section.

The state-to-state radiative transition probabilities $A_{ij}$ were obtained from experimental data when available \cite{Setser86,Alford92,NIST,Jung09,Nick85} but, fundamentally, from the theoretical survey of Aymar and Coulombe \cite{TheoDecay} and the recent theoretical compilation of Dasgupta \cite{Dasgupta}, that compare favourably with known values, at the level of 20\% or better. The width of the atomic transitions includes the natural width, the Doppler and collisional broadening, although for convenience a 1 nm measurement resolution has been assumed throughout this work. For resonant states strongly coupled to the ground state an effective increase of the state lifetime according to Holstein theory must be used \cite{Holstein}. The number of emission-absorption cycles for the cm-scale gas cells discussed in this work is $n_H\sim 1000$ \cite{SalameroNH}. In the pressure range studied here the practical contribution from such decays represents a \%-effect at most, in particular for the 1s$_4$ state and at the lowest pressure considered (0.1 bar).

Lastly, it is necessary to include a prescription on how the d and p multiplets are coupled, so that the cascade of the d-states can proceed down to the ground state within realistic times. Following \cite{Bruce90}, we assume that 2d$_5$ is fully coupled to 2p$_5$. A similar prescription is needed for the effective state Xe$^{**}$, that we couple directly to the highest atomic state explicitly considered (4d$_2$). Table \ref{TableI} lists the atomic reaction rates employed while table \ref{TableII} gives the matrix of the most relevant state-to-state probabilities.

From the above discussion and the numbers in table \ref{TableI} it becomes clear that discrete emissions from the atomic cascade are very unlikely above 1 bar in xenon, at the 1-2\% level and dominated by 2p$_{10-5}$ transitions (800-1100 nm). Their time scale is set to:
\beq
\tau^* \simeq \frac{\tau}{1+\tau K_2} = 0.1\tn{-}1 \tn{ns} ~~ \tn{(P}\gtrsim\tn{1 bar)}
\eeq
where $\tau^*$ indicates the effective life-time of a state, $\tau \equiv \tau_i = (\sum_j A_{ij})^{-1}$ represents its natural lifetime and $K_2$ its 2-body quenching rate ($\tau K_2 \simeq 100$-$1000$ above 1 bar). For particle detectors this observation qualitatively anticipates the dominance of excimer emissions over atomic ones, in the case of xenon.

\subsubsection{Excimer cascade}

The three most intense xenon `continuous' emissions in the range 100-1500 nm have been identified long ago and can be found described for instance in the works of Moutard \cite{Moutard}, Borghesani \cite{Borghe}, and references therein. We will refer to these emissions as $1^{\tn{st}}$, $2^{\tn{nd}}$ and IR continuum. Evidence for a UV continuum in the 300nm range has been found for $\alpha$ particles in \cite{Millet} and attributed to excimers formed starting from Xe$^{**}$ states and decaying to the dissociative A$1_g/0_g^-$ state.\footnote{We follow here the usual notation for the closely bound molecules (Hund's case $a$) and for the far bound molecules (Hund's case $c$). For details, see caption of table \ref{TableIII} and \cite{Demtroder}.} The effect has been included here for completeness, together with parameters from the authors. However the accuracy of the simulations depends critically on the extent to which the effective Xe$^{**}$ state assumed in that work and the one here behave similarly, therefore results in that region should be taken with care at the moment.

\begin{figure}[h!!!]
\centering
\includegraphics*[width=\linewidth]{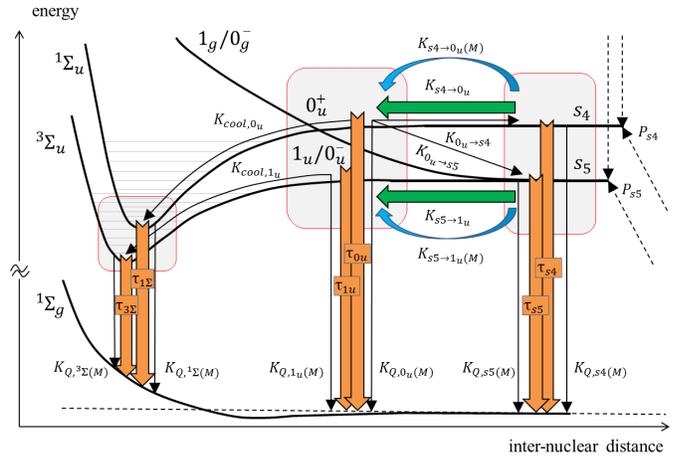}
\caption{Xenon pathways leading to the emission in the $1^{\tn{st}}$ and $2^{\tn{nd}}$ continuum as obtained in \cite{Moutard} for pure xenon (numerical values and additional references are given in table \ref{TableIII}). Bottom-up or top-down probabilities of populating the 1s$_4$ and 1s$_5$ states are indicated with dashed arrows. Additional pathways arising from the presence of additives are labeled with `(M)' (\cite{Wojche}, \cite{Setser78}). Potential curves and vibrational levels are artistic, scale is approximate. (The $X$ and $A$ letters usually prepended to the molecular states' labels, as well as the $n'$ index prepended to the atomic states, have been omitted in the figure for clarity.)}
\label{Decay2ndScheme}
\end{figure}

For the case of the VUV continua, the excimers relevant to the emission are known to be A$0^+_u$, A$1_u/0_u^{-}$ ($1^{\tn{st}}$ continuum) and the singlet and triplet $A\Sigma_u$ states ($2^{\tn{nd}}$ continuum). The generally accepted pathways are indicated in Fig. \ref{Decay2ndScheme} and can be found in pure xenon for instance in \cite{Moutard}, or more recently in \cite{Marchal}. Thin arrows have been used to indicate 2-body reactions and thick arrows indicate 3-body reactions as well as decays. To the standard pathways assumed for pure xenon there have been added the ones enabled in the presence of additives, in particular the 3-body quenching rates measured by Wojchechovski in \cite{Wojche}. They have been tagged with `$(M)$' and are introduced in the next sub-section. The probabilities to populate the 1s$_4$ and 1s$_5$ states either from cascade or electron impact are indicated by dashed arrows. They depend entirely on the seeding mechanism and can be thus obtained after performing the electron transport and computing the cascade. Numerical values for the parameters determining the xenon VUV-continuum as from Fig. \ref{Decay2ndScheme} are given in table \ref{TableIII}.

Based on detailed calculations, Borghesani et al. have recently suggested that the 2p$_5$ state could be a precursor of the IR continuum observed at around 1250nm both for the primary scintillation of 60keV electrons \cite{Borghe} and $\alpha$ particles \cite{IRalphas}. Although no dedicated experiment has been performed to assert this, the quantitative spectral agreement reported in \cite{Borghe} remains compelling. On the other hand, the high relative importance of 3-body reactions for the quenching of the 2p$_5$-state (table \ref{TableII}) hints at a mechanism analogous to that for the low-lying s-states being at play, with a third body stabilizing a newly formed excimer. Hence, as proposed in \cite{Borghe}, a diagram analogous to the one for the VUV continua has been assumed, and we take a lifetime of $\tau=1$ns for the 2p$_5$-associated excimers (decaying to the dissociative A$1_g/0_g^-$ state), neglecting quenching.\footnote{It is difficult to extract absolute yields (and thus assessing quenching) from electron gun measurements. According to \cite{Borghe2}, yields exhibit a transient behavior, dropping by about a factor x1/2 after 2h. This is presumably due to carbonaceous debris emitted from the entrance window. For data taken in reasonably short sequence the IR yield dropped by only 25\% when the pressure raised from 1 to 10 bar \cite{Borghe2}.} An effective excimer lifetime significantly larger than 1 ns for pressures above 2 bar would contradict indeed earlier measurements performed for 2p$_5$ selective excitation (see next section).

\begin{figure}[h!!!]
\centering
\includegraphics*[width=\linewidth]{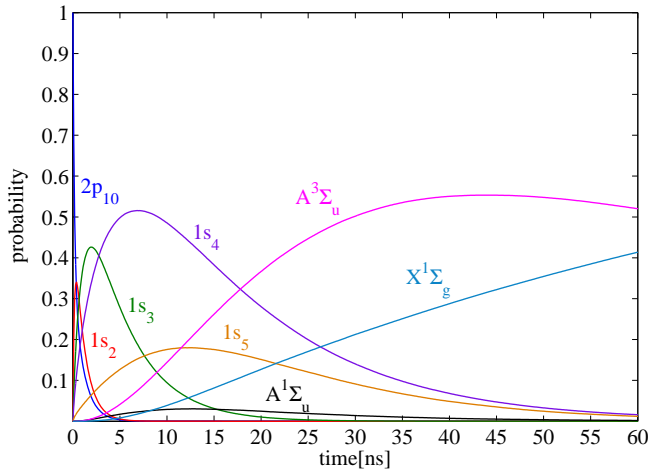}
\caption{Simulated cascade in pure xenon after selective excitation of the 2p$_{10}$ state at 1 bar. The intermediate $0^+_u$, $1_u/0_u^{-}$ and final $0_g^+$, 1s$_1$ states have very small relative populations at every instant of time and have been omitted for the sake of clarity. The singlet and triplet $A\Sigma_u$ states are the precursors of the $2^{\tn{nd}}$ continuum.}
\label{sel}
\end{figure}

The shape of the emission continua changes mildly with pressure by up to $10$-$20\%$ in the range 0.1-10 bar (\cite{Borghe}, \cite{Marchal}). Due to the difficulty at realizing a global model that includes this effect through a microscopic picture, and its a priori small technological importance, the continua have been parameterized by pressure-independent Gaussians. Their widths (full width half maximum) are taken to be 4 and 10 nm ($1^{\tn{st}}$ continuum), 12 nm ($2^{\tn{nd}}$ continuum), 100 nm (IR and UV continua), centered around 150nm, 170nm and 1250nm, respectively \cite{Marchal,Koehler,Borghe,Millet}. The evolution of the atomic and molecular cascade in pure xenon is illustrated in Fig. \ref{sel} for the case of selective seeding of the 2p$_{10}$ state.

\subsubsection{Molecular additives}

In pure xenon, the scintillation from 1s$_4$, 1s$_5$ and associated excimers is virtually unquenchable through collisions with xenon itself (Fig. \ref{Decay2ndScheme}), thus leading to a strong pressure-independence of the scintillation energy spectra above few 100's of mbar. However once a molecular additive is present, quenching rates usually at the scale of the solid sphere model are everywhere enabled, drastically changing the picture. Rates for 2-body collisions with additives can be found for low lying states mostly in \cite{Setser78}, and in \cite{Alekseev99, Alekseev96} for the high lying ones. By denoting as $f$ the additive concentration, one finds for instance 2-body quenching rates $K_2=11.1f, 12.8f, 18.8f$ [ns$^{-1}$] for the 1s$_5$, 1s$_4$ and 1s$_2$ states in case of CO$_2$ at 1 bar. Already from the $s_2$-state quenching rates and after inspection of table \ref{TableI} it becomes clear that, for the sub-\% additive admixtures studied in this work, the atomic cascade of high lying xenon excited states proceeds via radiative and collisional de-excitation with xenon itself. This is fortunate because the quenching rates of high lying states are also much less well known in the case of additives. They have been estimated here by analogy with similar molecules and states, for which the quenching of higher order p and d-multiplets have been measured, e.g., CH$_4$ and CF$_4$. Variations from molecule to molecule are typically within factors of $\times 2$-$3$, insufficient in any case to substantially modify the cascade dynamics for the sub-\% concentrations discussed here.

There is scarce information about the end-states in case of energy transfers to molecular additives, so we follow the conventional wisdom that the excitation energy is fully transferred, bringing the noble gas to the ground state \cite{Alekseev99}. The resulting CO$_2$ and CH$_4$ excited states and fragments are expected to be weak scintillators at typical operating conditions in gaseous detectors \cite{ArCO2Coimbra,CH4,Siegmund,Suzuki}, while for CF$_4$ the associated scintillation \cite{pureCF4} has been neglected due to the small concentrations discussed (at the level of 50-300ppm$_{_V}$). So, in general, it has been assumed as a proxy in simulations that additives do not lead to a sizeable re-emission in the studied range between 100 nm and 1500 nm. For the comparison with present data the quantum efficiency of the light-sensors used is actually limited to a much shorter range: 150-600nm at most.

Two last physical processes must be separately discussed:
\begin{enumerate}
\item Collisions between xenon atoms and additives via 3-body encounters, a process indirectly measured by Wojchechovski in a series of works performed on the 1s$_4$ state \cite{Wojche}. Although no detailed information about the end-state was given, we adopt as the default scenario the one qualitatively discussed there (as well as in \cite{Firestone}), where additives help at stabilizing the xenon excimer (Fig. \ref{Decay2ndScheme}, curved blue arrows). For high lying states, 3-body collisions with additives have been assumed to scale as the ratio of the 3-body/2-body collision rates measured in pure xenon, although this assumption has no influence on the results presented here:
    \beq
    K_{3(M)} = \frac{K_{3}}{K_{2}} K_{2(M)} \label{Rate3}
    \eeq
    Rate constants in eq. \ref{Rate3} refer to the same state.
\item Collisions between xenon excimers and additives. Since the rates for these processes are not known for xenon, we resort to the values measured for the associated atomic states, according to the observations made for instance for Ar$^*$ and Ar$_2^*$ in the case of CO$_2$, CH$_4$ and several other additives in \cite{ArExcQuenching}.
\end{enumerate}

The main uncertainties for a reliable prediction of the scintillation of doped xenon can be a priori expected from the last two processes (i.e., the nature of 3-body reactions with additives and the excimer quenching), so different extreme situations are evaluated in section \ref{Pdep}, and their agreement with present data discussed.

\section{Scintillation in pure xenon and comparison with data} \label{PureXe}

\subsection{Time spectra}

\begin{figure}[h!!!]
\centering
\includegraphics*[width=\linewidth]{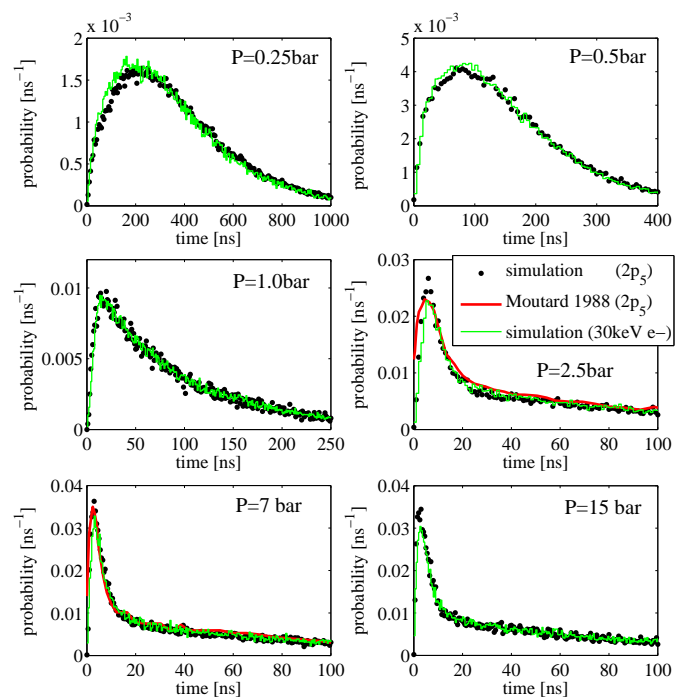}
\caption{Time spectra for the xenon $2^{\tn{nd}}$ continuum ($155\tn{nm}< \lambda <195\tn{nm}$) obtained at different pressures ($T=20^o$C) and for different seeding mechanisms. Dots show the results from simulations of the selective excitation process of the 2p$_5$ state, that is compared with data (\cite{Moutard}) at 2.5 and 7 bar (red thick lines). Results from the simulated scintillation spectra for 30 keV electrons are super-imposed for comparison (green thin lines).}
\label{Pscan}
\end{figure}

The time characteristics of the xenon 2$^{\tn{nd}}$ continuum have been benchmarked with data taken via selective excitation of the 1s$_2$, 2p$_9$ and 2p$_5$ states in the pressure range 0.1-15 bar in \cite{Moutard}, for reasons that will become clear soon. Although those measurements were made before the 2p$_5$ state had been conjectured as a precursor of the IR-continuum, they can be nowadays used to constrain some of the parameters involved: e.g., a reasonable agreement between data (thick red lines) and simulations (dots) in Fig. \ref{Pscan} requires the effective lifetime $\tau^*$ of the 2p$_5$-associated excimers to be below 2~ns for pressures above 2~bar.

\begin{figure}[h!!!]
\centering
\includegraphics*[width=\linewidth]{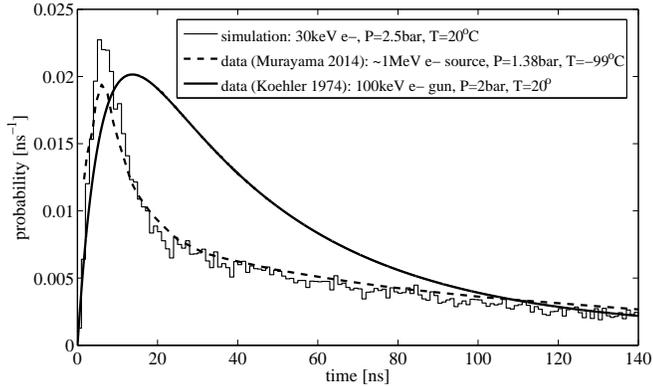}
\caption{Time profiles corresponding to the primary scintillation of electrons, obtained in data and in simulation ($155\tn{nm}< \lambda <195\tn{nm}$) for a xenon density around $N=5 \times 10^{19}$ cm$^{-3}$. Simulations are represented by thin lines and the recent measurements of Murayama and Nakamura \cite{Naka} are overlaid as dashed lines. The thick line shows the measurements by Koehler \cite{Koehler}, with an electron gun. In the experimental cases, the fitting functions provided by the authors have been used. It must be noted that, in simulation, the time profile is insensitive to the energy of the electron in the range given above.}
\label{timeComp25}
\end{figure}

It is important to note that, in simulation, the spectrum arising from the selective excitation of the 2p$_5$-state (dots) resembles very closely the one from 30 keV electron impact (thin lines) despite the former being, naturally, slightly faster. A similar situation can be observed for the electroluminescence spectrum (see for instance Fig. \ref{EL_30keV_spect}-up). This approximate universal behaviour supports our bench-marking case because the 2p$_5$ selective excitation experiment in \cite{Moutard} was made to be recombination-free, contrary to experiments dealing with intense electron guns, that can show an artificially slower (or faster) time evolution depending on the density conditions (e.g. \cite{Lorents}). The effect is illustrated through Fig. \ref{timeComp25}, where a comparison of different experimental situations at number densities around $N=5 \times 10^{19}$ cm$^{-3}$ is performed.\footnote{Measurements for $\alpha$-particles in similar density conditions have been performed in \cite{Suzuki}, displaying a spectrum compatible with the one shown in Fig. \ref{timeComp25}, except for the presence of recombination light at the 10's of ${\mu}s$ scale. }

The closeness between the temporal features observed for selective excitation of the 2p$_5$ state and for primary scintillation can be easily understood. Simulations show that, for energetic electrons (keV-MeV scale), around 50\% of the populated states are above 2p$_5$, and the characteristic time that it takes for the cascade to reach that atomic level is of the order of 1 ns above 1 bar. This is much shorter than the time constants of the singlet and triplet states, which dominate the scintillation process in that pressure regime ($\tau_{_{1\Sigma}}\!=\!4.5$ ns, $\tau_{_{3\Sigma}}\!=\!100$ ns). For pressures below 1 bar on the other hand the cascade becomes slower and quickly exceeds the shortest excimer time constant. The excimer formation rates become however the relevant time constants in those conditions, decreasing quadratically with pressure. The effect of the cascade duration is again diluted. Finally, provided the contribution of additional excimers above the 2p$_5$ level is small according to existing experimental evidence, it can be concluded that the time and spectral distributions for 2p$_5$ selective excitation and for the scintillation upon the impact of energetic electrons should be very similar.

\begin{figure}[h!!!]
\centering
\includegraphics*[width=\linewidth]{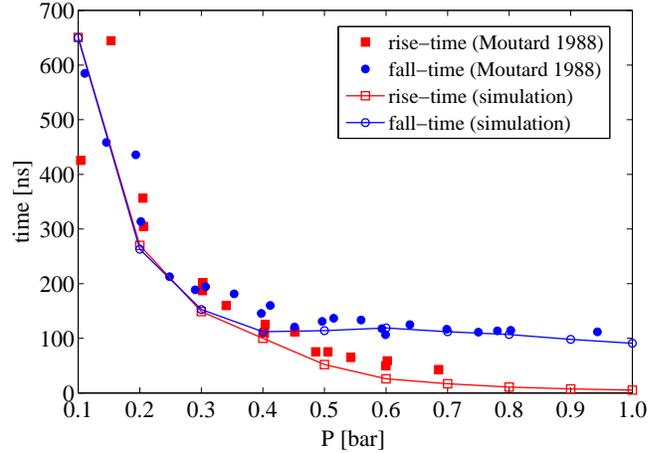}
\caption{Experimental (\cite{Moutard}) rise and fall-time constants of the xenon $2^{\tn{nd}}$ continuum below 1bar and comparison with simulation, for the case of selective excitation of the 1s$_2$, 2p$_9$ and 2p$_5$ states. Measurements are considered jointly in the experimental series. In simulation, no significant difference is observed for either case, so the 2p$_5$ is chosen for simplicity as the seed state.}
\label{tconstants}
\end{figure}

The above discussion on time constants can be better quantified by noting that below 0.8 bar the temporal spectrum is fitted well by a single component with corresponding rise and fall times (determined largely by the excimer formation rate and the lifetime of the triplet state), whereas spectra above 2 bar require of two components with two different fall times, each asymptotically approaching the life-time of the singlet and triplet states:
\bear
& \frac{dN_{\gamma}}{dt}\bigg|_{2^{\tn{nd}}} \simeq a e^{-t/t_{f}}- b e^{-t/t_{r}}, ~~~~ &(P\lesssim 0.8 ~ \tn{bar}) \\
& \frac{dN_{\gamma}}{dt}\bigg|_{2^{\tn{nd}}} \simeq a e^{-t/t_{f,fast}} + b e^{-t/t_{f,slow}}, ~~~~ &(P\gtrsim2.5 ~ \tn{bar})
\eear
with $a$, $b$ being defined positive. In the region around 1bar a fit with 8 time constants would be needed, becoming unstable. At high pressure on the other hand, the time spectra becomes a perfect double-exponential, with: $t_{f,fast}= \tau_{_{1\Sigma}}, t_{f,slow}= \tau_{_{3\Sigma}}$. Below 1 bar, the comparison between data and simulation can be more readily performed by plotting the $t_f$ and $t_r$ values resulting from the fit, as shown in Fig. \ref{tconstants}.\footnote{The spectrum reported for $P=0.5$ bar in \cite{Moutard} was found to be inconsistent with the $t_f$ and $t_r$ analysis reported by the same authors, as well as with the results obtained here, so it has been omitted. A good agreement is obtained if the pressure is assumed to be 0.4bar instead of the reported 0.5bar.}

The description of the time spectra below 1 bar is rather sensitive to the assumptions on the 3-body reactions $K_{s_4 \rightarrow 0_u}$ and $K_{s_5 \rightarrow 1_u}$ (Fig. \ref{Decay2ndScheme}). At the same time, while the value chosen for $K_{s_4 \rightarrow 0_u}$ is well within the commonly accepted values, $K_{s_5 \rightarrow 1_u}$ differs by a factor $\times 2$-$3$ relative to other estimates \cite{Marchal}, (table \ref{TableIII}). Indeed, a better description of the experimental spectra would require $K_{s_4 \rightarrow 0_u}$ to be about 20\% smaller than its presently assumed value. This discrepancy at low pressure is likely to be attributed to the complex spectral mixing between the $1^{\tn{st}}$ and $2^{\tn{nd}}$ continuum taking place precisely in the wavelength region where the quartz window used in \cite{Moutard} has a cutoff \cite{Marchal}. We plan to revise these assumptions more critically in the future, but we keep them here in order to mantain the phenomenological value of the simulations. On the other hand, for pressures of 1 bar or above, the assumptions on $K_{s_4 \rightarrow 0_u}$ and $K_{s_5 \rightarrow 1_u}$ have an impact on the light yield in the $2^{\tn{nd}}$ continuum below 5\%.

\subsection{Yields}

\begin{figure}[h!!!]
\centering
\includegraphics*[width=\linewidth]{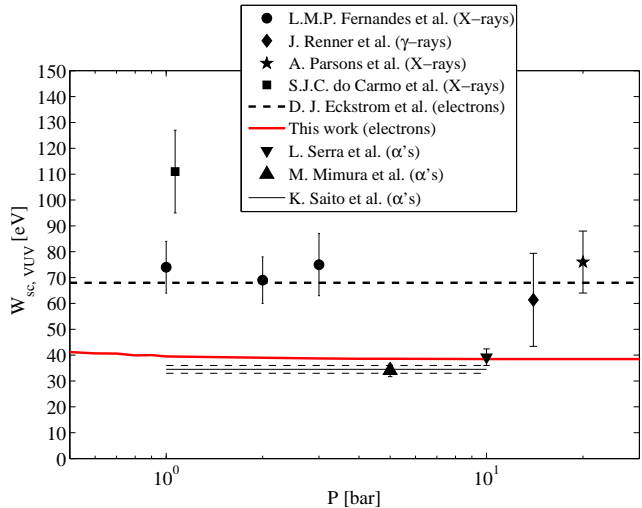}
\caption{World-compilation of $W_{sc,VUV}$ in pure xenon (average energy to produce a VUV photon) in the absence of charge recombination. Data corresponds to Fernandes \cite{Pancho} (circles), Renner (this collaboration) \cite{Josh} (diamond), Parsons \cite{Parsons} (star), do Carmo \cite{Filomena} (square), Serra (this collaboration) \cite{Serra} (down-triangle),
Mimura \cite{Mimura} (up-triangle) and Saito \cite{Saito1} (thin lines). The calculation of Eckstrom for electrons \cite{Eckstrom} is indicated by a thick dashed line, whereas the thick line represents the simulation described in this work. $T\simeq20^o$C in all cases.}
\label{Wvalue}
\end{figure}

Information about xenon scintillation yields of individual charged particles in the absence of recombination is scarce, and values for the average energy to create a photon vary in the range $W_{sc}=30$-$120$eV (Fig. \ref{Wvalue}). Most of the measurements in the literature refer to the VUV region, except for the IR measurements performed in \cite{IRalphas}. As shown in \cite{Serra}, recombination light amounts at most to 10\% of the total scintillation light observed for $\alpha$ particles at 10 bar, as long as $E/P > 30$ V/cm/bar. For electrons and X-rays the effect is even smaller \cite{Balan}. In those conditions, the average energy to create a VUV photon in xenon depends just on the population of excited states, $N_{ex}$, that are directly promoted by the impinging particle (of energy $\varepsilon$):
\beq
W_{sc,VUV} \equiv \frac{\varepsilon}{N_{\gamma}} \simeq \frac{\varepsilon}{N_{ex}}
\eeq
This relation is exact in simulation, stemming from the fact that any excited state cascades down to one of the VUV-precursors, and so the number of photons, $N_{\gamma}$, satisfies $N_{\gamma}/N_{ex}=1$ in pure xenon.
A world-compilation of $W_{sc,VUV}$ values obtained under fields at the scale of 30 V/cm/bar or higher is given in Fig. \ref{Wvalue}, together with the results from the simulation described in text (thick line). Dashed lines show the calculation of Eckstrom et al. \cite{Eckstrom}. For higher realism, simulations consider emission in the second continuum window ($155\tn{nm}< \lambda <195\tn{nm}$), that leads to a small increase of $W_{sc,VUV}$ at low pressure due to competition from the $1^{\tn{st}}$ continuum.

It is clear from Fig. \ref{Wvalue} that values for $W_{sc,VUV}$ tend to be systematically higher for $X$-rays, $\gamma$-rays and electrons (nearly a factor 2) compared to values measured for $\alpha$ particles. As experimentally observed, the simulation results depend just mildly on pressure, and they indicate as well a very similar response to either X-rays, $\gamma$-rays or electrons. Notwithstanding this, the region of $W_{sc}$-values corresponding to $\alpha$ particles ($\sim40$eV) is clearly favoured in simulation, in detriment of the lepton band ($\sim70$eV).

It should be noted that measurements done with X-rays rely on a very faint light signal, and special analysis techniques are needed, all suffering from the absence of a clear energy peak. Measurements performed by this collaboration in \cite{Josh} are an exception, due to the high $\gamma$-ray energy used, however the relatively large error bars do not allow an unambiguous  conclusion (diamond). At the same time, the normalization of the secondary scintillation is reproduced in simulation within a 5-10\% accuracy (see next section), so if the larger experimental values are assumed to be correct, the discrepancy between data and simulations should be attributed to a deficient high energy extrapolation of the elementary cross-sections used in Degrad. At any rate, future theoretical and experimental work should aim at clarifying the difference between electron and $\alpha$ particle excitation, in order to either explain or exclude this deviation.

The present simulation code can be used to obtain a value for the near infra-red scintillation ($700\tn{nm}< \lambda <1500\tn{nm}$), yielding $W_{sc,IR}=86$ eV at 2.5bar, to be compared with the measured value $W_{sc,IR}\lesssim48\pm 7$ eV obtained in \cite{IRalphas} for $\alpha$ particles.

\section{Scintillation in doped xenon and comparison with data} \label{MixXe}

To the authors knowledge, there is no systematic study of the scintillation time constants for xenon gas in the presence of additives, namely, obtained under conditions typical of particle detectors' operation. We will concentrate therefore on the overall light yield and its dependence with field and pressure, together with the fluctuations of the scintillation process. The time spectrum is later shown as a prediction (section \ref{Pdep}).

\subsection{Yields} \label{Yields}

Secondary scintillation for pure xenon and for xenon doped with CO$_2$ has been measured recently at around 1 bar in \cite{Joaquim0} and \cite{Joaquim1}, respectively, while the full pressure range up to 10bar was covered in \cite{Joaquim2}. %Equivalent measurements for CH$_4$ and CF$_4$ are being performed by this collaboration and will be presented elsewhere, however they do not alter the general conclusions found here.
Concentrations were calibrated and monitored with the help of a residual gas analyzer (RGA), their uncertainty found to be at the 100 ppm$_{_V}$ level, typically. We present here as well data taken with the NEXT-DBDM demonstrator \cite{DBDM} for CF$_4$ admixtures at 10 bar, both for primary and secondary scintillation. This gives an impression about the ability of the present simulations to extrapolate to high pressure as well as to the primary scintillation. Such data were taken without RGA, by relying on pre-mixed amounts. Unlike CO$_2$, we have observed that CF$_4$ shows generally a small reactivity with cold gas getters and, besides, the swarm characteristics were found to be in agreement with the expectation from simulations (Fig. \ref{vdDl}).

The optimal concentration necessary to reduce diffusion in xenon from 10~mm to the mm-scale required to significantly enhance pattern-recognition in a $\beta\beta0\nu$ experiment is found, with the help of Magboltz, to be in the range of admixtures of $0.05$-$0.1\%$ for CO$_2$ and $0.01$-$0.02\%$ for CF$_4$. With such concentrations, and in the range of drift fields E/P = 20-30 V/cm/bar, the characteristic size of the charge-diffusion ellipsoid is estimated to be:
\beq
\sigma_{_{3D}} =\sqrt[3]{\sigma_{x}\sigma_{y}\sigma_{z}} \simeq 2\tn{-}2.5 \tn{~mm}\sqrt{\frac{10~\tn{bar}}{P}} ~~ \tn{(after 1~m drift)} \label{TPC-diff}
\eeq

\begin{figure}[h!!!]
\centering
\includegraphics*[width=\linewidth]{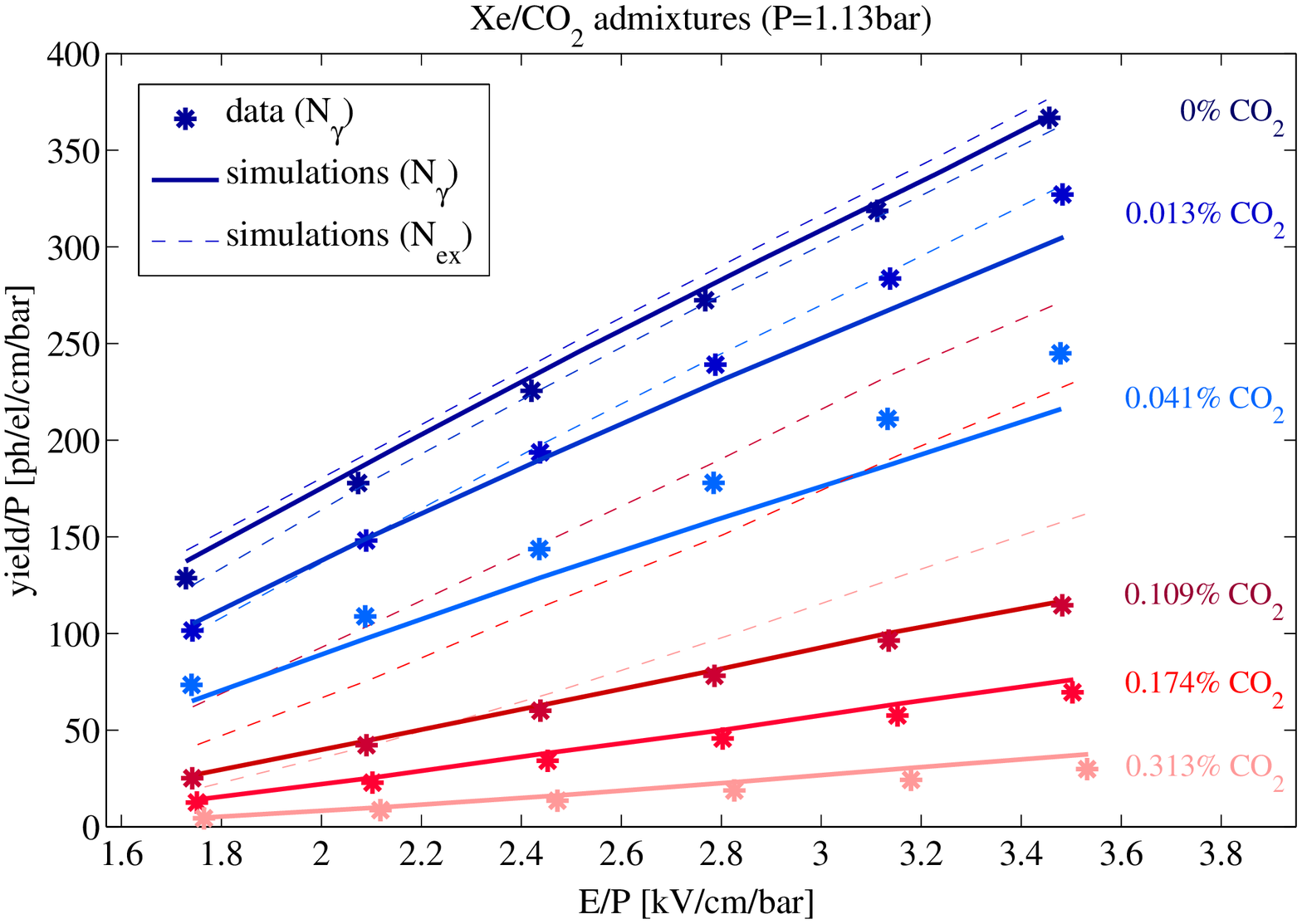}
\includegraphics*[width=\linewidth]{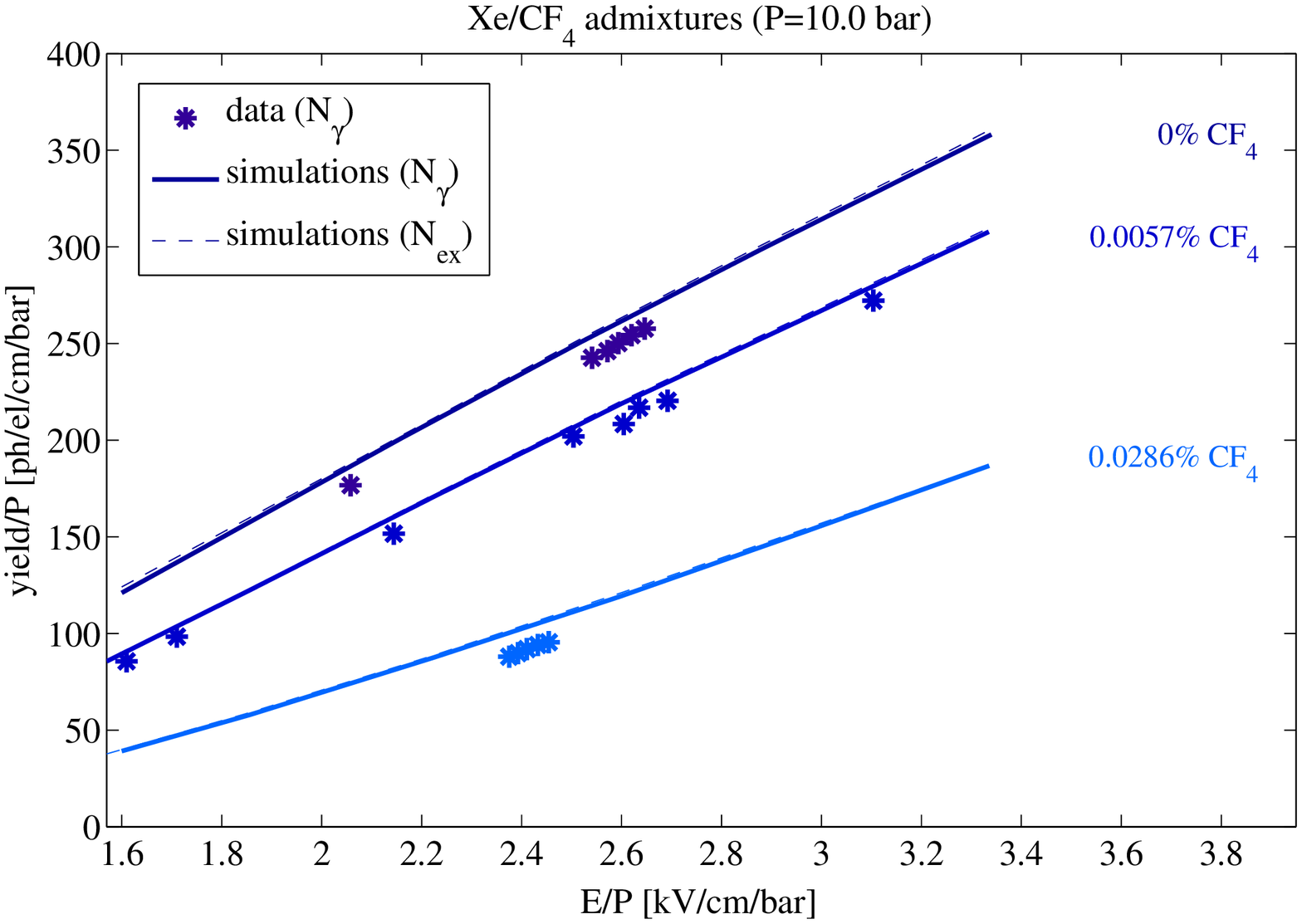}
\caption{Top: secondary scintillation data in the 150-600nm range for CO$_2$ admixtures after \cite{Joaquim1} (asterisks), simulated number of photons, $N_\gamma$ (continuous lines), and simulated number of excited states, $N_{ex}$ (dashed lines). The chamber pressure was 1.13bar and the gas gap 25mm. Bottom: secondary scintillation data in the 150-600nm range for CF$_4$ admixtures as obtained in NEXT-DBDM detector. The chamber pressure was 10.0bar and the gas gap 5mm. Absolute normalization for pure xenon data assumed to be identical to the 1bar case. The size of the data points is bigger than their statistical uncertainty.}
\label{yield_}
\end{figure}

Present efforts focus therefore on those concentrations. A comparison between experimental data and simulations in the wavelength range 150-600nm for both CO$_2$ and CF$_4$ mixtures is given in Figs. \ref{yield_}, in which the secondary scintillation yields are plotted for various additive concentrations and fields. A general agreement is found in the CO$_2$ case, deviating up to 30\% for the highest concentration, while CF$_4$ yields are systematically over-estimated in simulation by about 10\%. For a deeper understanding of the situation, it is useful at this point to separate the effects coming from electron transport from those coming from the scintillation process, provided they are fully uncorrelated. We studied for that the ratio of photons in the $2^{\tn{nd}}$ continuum ($N_{\gamma}$) relative to the number of excited states obtained in simulation ($N_{ex}$), a ratio that we conveniently refer to as the `scintillation probability', $\mathcal{P}_{scin}$. They are represented by continuous and dashed lines in Figs. \ref{yield_}, respectively. The scintillation probability is complementary to the quenching probability for VUV-scintillation ($P_Q$):

\beq
 \mathcal{P}_{scin}  = 1-\mathcal{P}_Q = \frac{N_{\gamma}}{N_{ex}} \label{Pscin1}\\
%& \mathcal{P}_{scin}|_{sim}  & = 1-P_q|_{sim}  = \frac{N_{\gamma}|_{sim}}{N_{ex}} \label{Pscin2}
\eeq

For pressures above 1 bar, scintillation in the $2^{\tn{nd}}$ continuum follows in simulation from any primary excited state, with a probability of 97\% or more,
thus $\mathcal{P}_{scin}\geq0.97$ for pure xenon. %The sub-indexes in eqs. \ref{Pscin1}, refer to whether the information is retrieved from data or from simulations.
Quite naturally, the value for  $\mathcal{P}_{scin}$ in the presence of additives must depend only on the scintillation process, as long as a reasonable estimate exists for $N_{ex}$. That value can be obtained directly from transport. The numerator $N_{\gamma}$ on the other hand can be evaluated both from data and simulation. To simplify the discussion it should be noted that, provided 1s$_4$ and 1s$_5$ populations depend little on the field, and the scintillation mechanism is similar for either atomic state, $\mathcal{P}_{scin}$ is nearly field-independent in simulation. Moreover, as previously argued, the rather fast nature of the cascade process makes any higher lying Xe$^*$ state rather immune to the presence of additives, so $\mathcal{P}_{scin}$ proves a useful measure of the available scintillation, either primary or secondary.

\begin{figure}[h!!!]
\centering
\includegraphics*[width=\linewidth]{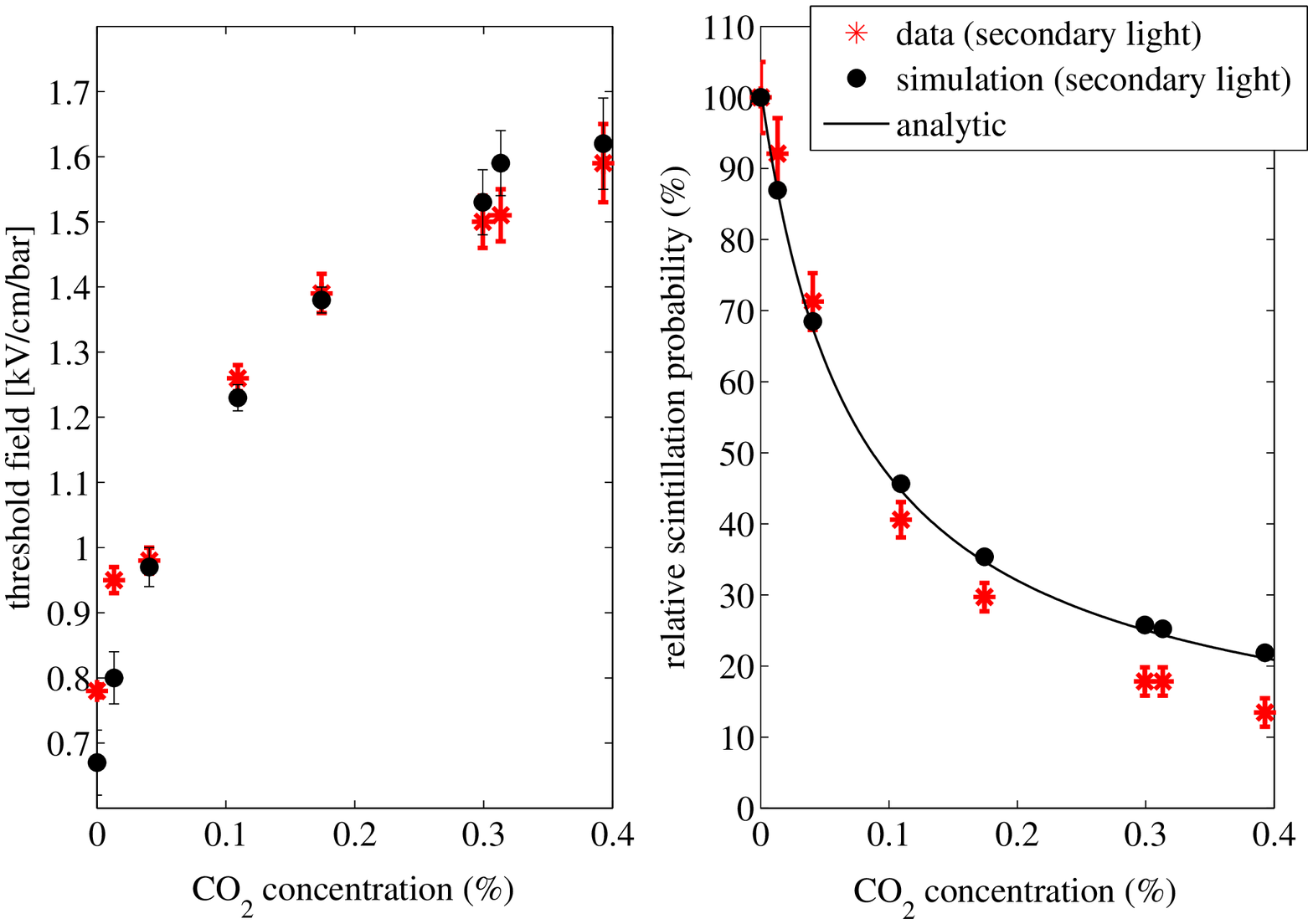}
\includegraphics*[width=\linewidth]{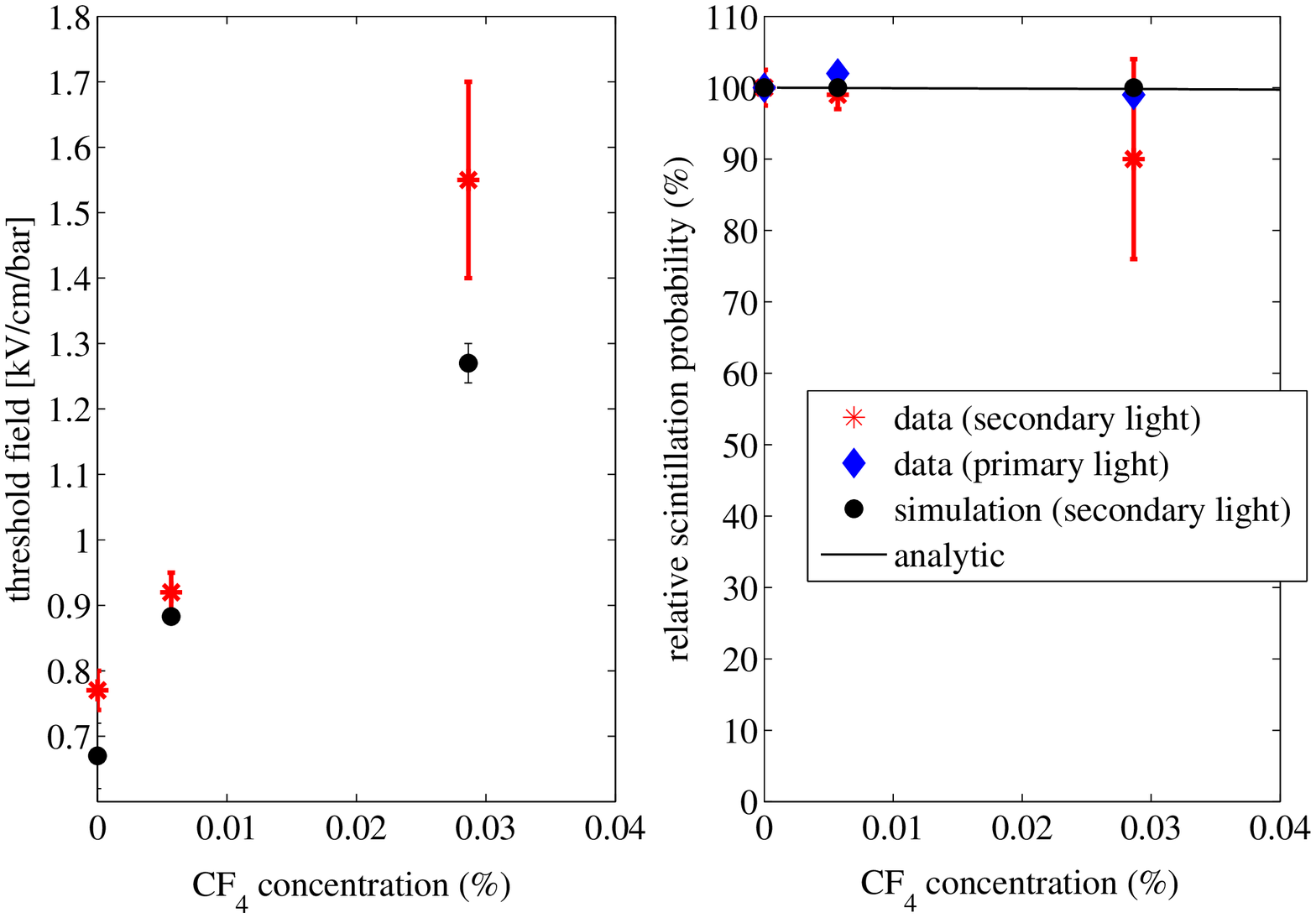}
\caption{Experimental and simulated parameters characterizing the primary and secondary scintillation in xenon mixtures. Left: threshold field, E$_{th}$. Right: scintillation probability, $\mathcal{P}_{scin}=N_{\gamma}/N_{ex}$, relative to pure xenon. The top figures are extracted from Xe-CO$_2$ mixtures at 1.13bar (25mm gas gap), and the bottom ones correspond to Xe-CF$_4$ at 10.0bar (5mm gas gap).}
\label{yield_th}
\end{figure}

But how well can $N_{ex}$ be known and $\mathcal{P}_{scin}$ reliably estimated from data?. An indirect assessment of the quality of transport can be extracted by realizing that electroluminescence is a linear process, and this behaviour is not strongly altered in the presence of additives. By fitting the observed trends in Fig. \ref{yield_} to straight lines it is possible to obtain the minimum field required for the gas to scintillate, $E_{th}$, that corresponds to the situation for which the electron characteristic energy is such that 1s$_4$ and 1s$_5$ promotion becomes sufficiently likely. The additional electron cooling introduced by the presence of the additive shifts the threshold field to higher values compared to pure xenon, in good agreement with simulation (Figs. \ref{yield_th}-left). The strongest effect takes place both in experiment and in simulation precisely for the concentration ranges where the cooling is expected to produce a sizeable suppression of the TPC diffusion according to relation \ref{TPC-diff}.

Complementary to $E_{th}$, the scintillation probability is given on Figs. \ref{yield_th}-right. $\mathcal{P}_{scin}$ is obtained directly from the ratio of the slopes of the $N_{ex}$ vs $E/P$ and $N_{\gamma}$ vs $E/P$ linear trends. It has been normalized to the scintillation probability in pure xenon (re-labeled as $\mathcal{P}_{scin}^*$), that is hence defined as 1 by convention. This normalization should not perturb the reader, as it represents a 3\% correction at most, however it helps at representation since it factors out a small discrepancy at the 5\% level between the slopes observed for pure xenon in data and simulation. This discrepancy is simply added to the experimental error bars. As shown later, $\mathcal{P}_{scin}$ allows for a convenient analytical expression, that compares accurately with the results from the full simulation (Fig. \ref{yield_th}-right, continuous lines).

The experimental behaviour of the scintillation probability for Xe-CO$_2$ mixtures shown in Fig. \ref{yield_th} clearly follows a simple quenching relation as a function of the additive concentration $f$, like $\sim\!1/(1 + \tau K_2 f)$. A direct fit, if naively assuming that $\tau=\tau_{_{3\Sigma}}$, yields a two body quenching rate $K_2=12.7 \pm 1.1$ ns$^{-1}$ ($K_2=11.2 \pm 1.0$ ns$^{-1}$ at 1bar), in excellent agreement with the expected value for the xenon triplet state: $K_{Q,^3\Sigma(M)}=11.12$ ns$^{-1}$ (table \ref{TableIII}). A detailed analysis of the contributions (see next section) shows that this is indeed the dominant process near atmospheric pressure, as intuitively expected. This is also the case for Xe-CF$_4$ mixtures, however the anomalous quenching rate of this molecule ($K_{Q,^3\Sigma(M)}=0.074$ ns$^{-1}$ at 10bar, \cite{Setser78}) renders a nearly flat behaviour, with a product $\tau_{_{3\Sigma}} \times K_{Q,^3\Sigma(M)} \times f$ extrapolating to a 0.3\% scintillation drop in the range displayed in Fig. \ref{yield_th}. This observation is consistent, too, with the absence of any measurable effect on the primary scintillation yields in the same conditions (diamonds).

\subsection{Light fluctuations}

The fluctuations of the light production process can be conveniently described by studying the relative standard deviation squared, that we name `$Q$-factor' \cite{Escada, homeo}:
\beq
Q = \frac{\sigma_{m_\gamma}^2}{m_\gamma^2}
\eeq
where $m_\gamma$ refers to the number of photons per primary electron (equivalently, the optical gain).
$Q$ can be decomposed in 3 main contributions: the fluctuations in the number of excited states that are precursors to scintillation, the fluctuations in the quenching process, and the fluctuations due to electron losses (e.g., attachment):
\beq
Q = Q_{ex} + Q_{\mathcal{P}_{scin}} + Q_{att} \label{Qdecompose}
\eeq
Once the intrinsic calorimetric response of a TPC needs to be evaluated for contained events, $Q$ adds linearly to the Fano factor $F$, in similar fashion to the multiplication statistics of a charge-amplification device $f_g$ \cite{RobFluc, Elisa}. The main difference is that, while $f_g$ is within 3-6 times bigger than the Fano factor in the case of xenon mixtures, $Q$ can be just 10\% of it, or even less. Because of that, the experimental extraction of $Q$ has been rather elusive so far. The last term in eq. \ref{Qdecompose} can achieve however measurable values in the presence of additives albeit, as demonstrated in \cite{Joaquim1}, with a fairly large experimental uncertainty (Fig. \ref{Q_CO2}). As shown in appendix \ref{attachSection}, this term is well approximated in the limit of small attachment by a simple analytical formula, whereas $Q_{scin}$ admits a direct derivation from the associated binomial probability distribution:
\bear
&Q_{att}&  \simeq \frac{1}{3} \eta g \\
&Q_{\mathcal{P}_{scin}}& = \frac{1}{m_\gamma}\mathcal{P}_{scin}(1-\mathcal{P}_{scin})
\eear
The attachment coefficient has been indicated as $\eta$ and $g$ is the electroluminescence gap. The contribution from $Q_{\mathcal{P}_{scin}}$ is generally negligible for high light yields, as is the case here. The comparison with data shown in Fig. \ref{Q_CO2} indicates that, indeed, at the fields characteristic of secondary scintillation, the dissociative attachment of CO$_2$ has a central role in the $Q$-factor of the admixtures, thus increasing with concentration in an approximate linear fashion. This is not at all the case for CH$_4$, for instance, as it will be shown elsewhere. The analytical expression for $Q_{att}$ with $\eta$ evaluated directly from Magboltz is overlaid in Fig. \ref{Q_CO2} (green lines).

\begin{figure}[h!!!]
\centering
\includegraphics*[width=\linewidth]{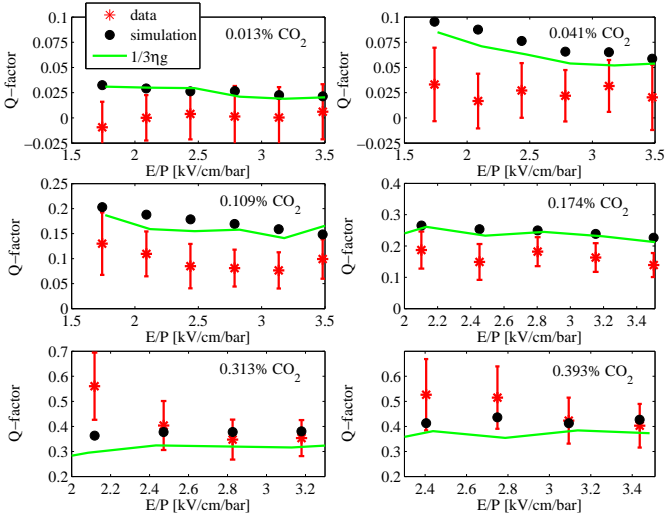}

\caption{$Q$-factor in Xe-CO$_2$ mixtures (asterisks) and comparison with simulation (circles). The approximate analytical expression for the case when $Q$ is attachment-dominated is indicated with continuous lines.}
\label{Q_CO2}
\end{figure}

It is worth noting that the impact of electron attachment on the light yield produces a relatively mild dependence as $m_\gamma \rightarrow (1-\frac{1}{2}\eta g) m_\gamma$ (appendix \ref{attachSection}), so a $Q$-factor exceeding $F$ by a factor as high as $\times3$ would in turn reduce the light yield in a very modest $\sim60\%$. Clearly, in the case of Xe-CO$_2$ admixtures the main cause of light loss is quenching (Fig. \ref{yield_th} up-right), and the main source of light fluctuations is dissociative attachment (Fig. \ref{Q_CO2}).

\section{Pressure dependence and model uncertainties} \label{Pdep}

\subsection{High pressure behavior}  \label{Pdependence}
The scintillation probability ($\mathcal{P}_{scin}$) for the xenon second continuum, following the pathway diagram in Fig. \ref{Decay2ndScheme} has an analytical solution as (Appendix \ref{analyticalPathways}):
\bear
&&\mathcal{P}_{scin, s_4}\!=\!\frac{\mathcal{P}_{pop,s_4} \cdot \mathcal{P}_{s_4\rightarrow{0_u}} }{1-\mathcal{P}_{s_4\rightarrow 0_u} \cdot \mathcal{P}_{0_u\rightarrow s_4 }} \times \label{ana1}  \nonumber \\
&& ~\left( \mathcal{P}_{cool,0_u} \!\!\cdot\! \mathcal{P}_{rad,^1\Sigma} + \mathcal{P}_{0_u\rightarrow{s_5}} \!\!\cdot\! \mathcal{P}_{s_5\rightarrow{1_u}} \!\!\cdot\! \mathcal{P}_{cool,1_u} \!\!\cdot\! \mathcal{P}_{rad,^3\Sigma} \right) \\
&&\mathcal{P}_{scin, s_5}\!=\!\mathcal{P}_{pop, s_5} \!\cdot\! \mathcal{P}_{s_5\rightarrow{1_u}} \!\cdot\! \mathcal{P}_{cool,1_u} \!\cdot\! \mathcal{P}_{rad,^3\Sigma} \label{ana2}
\eear
with $\mathcal{P}_{scin} = \mathcal{P}_{scin, s_4} + \mathcal{P}_{scin, s_5}$. Reading the second equation from left-right we find the probability of populating the atomic state, the probability that the atomic state ends forming a far-bound excimer, the probability that it forms a close-bound excimer, and the probability that such an excimer decays. The first line contains an additional pre-factor to account for the probability that the far-bound excimer is dissociated.
These probabilities can be exactly computed starting from the reaction rates given in Fig. \ref{Decay2ndScheme} and tabulated in table \ref{TableIII}, and have been plotted in Fig. \ref{AnalyticPathways} (see Appendix \ref{analyticalPathways}). A convenient approximation can be obtained by observing that all but the radiative probabilities are little dependent on the additive concentration for the ranges of practical interest (sub-$\%$), except for a small 10\% effect. By further adopting $\mathcal{P}_{pop,s_4} \sim \mathcal{P}_{pop,s_5} = 0.5$ (a reasonable approximation according to simulation) the scintillation probability reads:
\beq
\mathcal{P}_{scin} \simeq \frac{F_1}{1 + f\!\cdot\! n\!\cdot\!\tau_{_{^1\Sigma}}\!\cdot\! K_{Q,^1\Sigma}} + \frac{F_3}{1 + f\!\cdot\! n\!\cdot\!\tau_{_{^3\Sigma}}\!\cdot\! K_{Q,^3\Sigma}} \label{AnaSimple}
\eeq
with $f$ being as before the additive concentration, and $n=P/P_o$ the pressure relative to $P_o=1$ bar ($T=20^\circ$C). The constants can be shown to be $F_1\simeq0.1$ and $F_3\simeq0.9$. It is thus not surprising that the triplet state dominates the scintillation properties, provided it is both slower and more efficiently populated than its singlet counterpart. This fact makes good the expression `triplet dominance model' (TDM) for referring to the experimental situation around atmospheric pressure. At 10 bar the singlet and triplet contribution become comparable, anticipating an additional $\times 2.5$ decrease of the light yield for CO$_2$ mixtures in the region of interest for NEXT, compared to 1bar.

\begin{figure}[h!!!]
\centering
\includegraphics*[width=\linewidth]{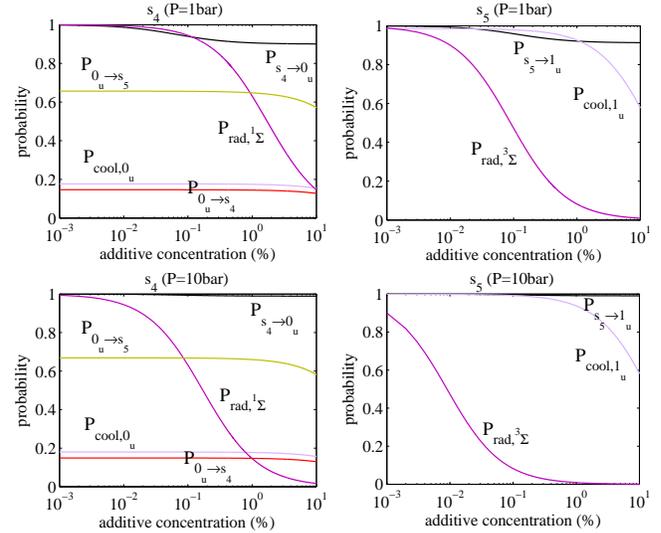}

\caption{Different terms contributing to the scintillation probability of electrons in xenon, as a function of the additive concentration, according to eqs. \ref{ana1}, \ref{ana2}. The radiative terms $\mathcal{P}_{rad,^{1(3)}\Sigma}$ dominate completely the scintillation behaviour in the sub-\% range.}
\label{AnalyticPathways}
\end{figure}

\begin{figure}[h!!!]
\centering
\includegraphics*[width=8cm]{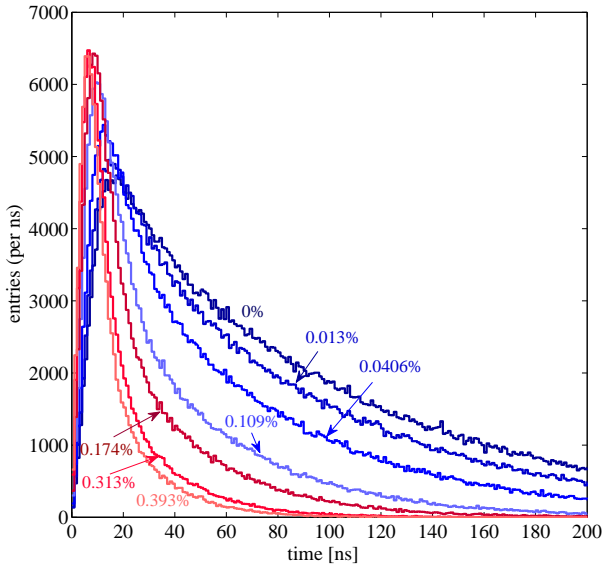}

\caption{Simulated time spectra of the xenon 2$^{\tn{nd}}$ continuum for different concentrations of CO$_2$ at $P=1.13$bar, and $T=20^\circ$C.}
\label{YieldAll}
\end{figure}

When looking at the simulated time characteristics of the scintillation (Fig. \ref {YieldAll}) it can be seen how the contribution from the tail of the triplet state virtually disappears at high additive concentrations, while the rise-time becomes faster, due to the assistance of 3-body collisions with the additive.

Light fluctuations show a dependence with pressure, as well, given by:
\beq
Q_{att}  \simeq \frac{n}{3} \eta(E/p) g
\eeq
an expression that inherits the $P$-scaling of the attachment coefficient in Magboltz. Since the gap used for the CO$_2$ measurements conveyed here is 5 times bigger than the canonical $\sim5$mm gap of the NEXT experiment, the $Q$-factor will nearly double at 10bar relative to the value plotted in Fig. \ref{Q_CO2}. For concentrations in the range $0.05-0.1$\% this implies doubling the Fano factor, at most.

\subsection{Model ambiguities and uncertainties}

\begin{figure}[h!!!]
\centering
\includegraphics*[width=8cm]{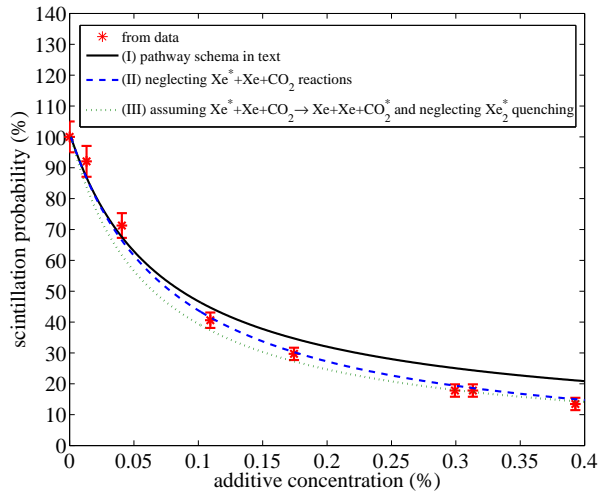}
\caption{Scintillation probability and comparison with the results of text (continuous lines), model II (dashed lines) and model III (dotted lines).}
\label{TH_SLOPE_MODELS}
\end{figure}

Although the proposed simulation framework makes use of sensible assumptions, the end-states of the 3-body collisions with additives as well as the quenching rates of the excimer remain unknown. We have relaxed those assumptions and studied several extreme possibilities by combining the following approximations, namely: i) 3-body collisions with additives leading to full atomic quenching instead of excimer formation, ii) absence of excimer quenching and iii) 3-body collision with additives being much smaller than previously measured. Surprisingly, two such models describe Xe-CO$_2$ scintillation data better than the default model used in text (Model I):
\begin{itemize}
\item Model II: 3-body collisions with additive are neglected.
\item Model III: 3-body collisions with additive bring the atom to ground state but excimer quenching can be neglected (the scenario previously assumed in \cite{Escada}).
\end{itemize}
Both are shown in Fig. \ref{TH_SLOPE_MODELS}, together with the default model used in text. Other combinations of assumptions i), ii) and iii) lead to either too high or too low scintillation. We describe briefly the implications.

Model III is certainly plausible, but it does not agree with earlier observations of 3-body reactions with additives accelerating excimer formation as well as the fact that atomic quenching and excimer quenching are similar for the case of argon mixtures, and it would be surprising if the latter will just become negligible for xenon. Interestingly, due to the absence of excimer quenching, the model exhibits nearly no dependence with pressure, and the time spectrum keeps its shape (contrary to Fig.\ref{YieldAll}). Therefore in the presence of some (even coarse) timing information this model could be easily validated/rejected experimentally. There are two additional shortcomings of model III: it is not capable of describing CH$_4$ scintillation data (that will be presented elsewhere), and it ignores the fact that the experimental scintillation yields can be described by a simple quenching relation (Occam's razor).

Much more compelling is the fact that a perfect agreement is found if 3-body collisions with the additive are neglected (model III). It turns out that such collisions are also not necessary to describe CF$_4$ and CH$_4$ scintillation data (and, even if included, they represent a small correction). The phenomenological power behind this observation is evident: if 3-body reactions with the additive are in general not essential to describe the quenching mechanisms in xenon-based mixtures, the simple approximation sketched in eq. \ref{AnaSimple} would allow to describe their scintillation characteristics, by relying on parameters that are often well measured in literature. The predictive power would then depend on the extent to which the atomic and excimer quenching rates are similar, something that has been verified for a number of Ar mixtures. Given the slightly different dissociation energies of atoms and excimers, such a behaviour cannot be expected to be completely universal, though.

As a conclusion, an experimental compromise seems to exist for a $\times (10\tn{mm}/2.5\tn{mm})^3$ reduced diffusion at 10bar in Xe-CO$_2$ and Xe-CF$_4$ mixtures as compared to pure xenon, while keeping the scintillation levels at about $\sim\!1/5$ relative to those of pure xenon, in the worst case. For CO$_2$-mixtures, the $Q$-factor stemming from the intrinsic fluctuations in the scintillation process will at most double the Fano factor. Based on the NEXT geometry and PMTs response this improvement is encouraging, with clear potential to keep the energy resolution at the level of $0.5\%$ at $Q_{\beta\beta}$, if it is not dominated by systematic effects. A comprehensive work on these and other practical considerations will follow once the experimental campaign is finished.

\section{Discussion: TDM on the light of previous results}

Systematic measurements of xenon scintillation in the presence of additives, for conditions of practical relevance in gaseous detectors, are scarce. Apparently a turning point was the realization by Policarpo, Conde and Alves, very early on, that a sizeable increase of the electron drift velocity was hardly compatible with maintaining a high light yield, except perhaps for nitrogen admixtures \cite{PolicarpoN2}.
If interpreting the observed light yield drop at constant field as dominated by light quenching alone, a quenching factor $P_Q\simeq0.5$ was obtained then for N$_2$ concentrations around 2\%. The theoretical value under the triplet dominance model,
and taking the 2-body quenching rate from \cite{Setser78} ($K_2=0.47$ ns$^{-1}$ at 1bar) is remarkably close, 2.1\%. A much stronger (and not understood) quenching effect was observed later by Takahashi et al. in \cite{Suzuki} and so in this way one of the best additive candidates (to that date) seems to finally have dropped.%\footnote{In fact Xe-N$_2$, according to Magboltz simulations, does not cool the electrons enough to be of use in gaseous optical TPCs even if assuming that TDM applies.}
The scintillation quenching measured in Ar-CH$_4$ and Ar-CO$_2$ by the same group was on the other hand shown to be fully compatible with TDM, while Ar-N$_2$ displayed a strong re-emission in the 300nm region \cite{Poli67}, thereby requiring a different theoretical framework. As confirmed by present work, the $\times 50$ higher quenching power observed for CO$_2$ in Ar relative to Xe can be now linked directly to the ratio of the triplet lifetimes (3200/100), with quenching rate itself amounting to the remaining factor of $\simeq2$.

Although high concentrations of CF$_4$ will eventually increase dissociative attachment to unbearable levels for any counter relying on secondary scintillation, the Berna group recently suggested in a series of works the possibility of using CF$_4$ as an additive in order to keep the primary scintillation and some compatibility with a charge amplification device \cite{Vulle1, Vulle2}. They reported `no degradation of the primary scintillation' for a Xe/CF$_4$(98/2) admixture at 3bar, that is again compatible with TDM, since the scintillation from the triplet state would be quenched down by just $5\%$ in those conditions ($K=0.0074$ ns$^{-1}$ at 1bar).

Measurements for the primary scintillation of xenon in the presence of CH$_4$ were been performed by Pushkin under $\alpha$-particles at around 26bar in \cite{Pushkin} but, showing a sizeable contribution from recombination light, they do not allow for a direct comparison. The CH$_4$ case will be discussed in detail in an ongoing publication where new data will be presented.

At last, although indirectly, some light emission characteristics can be inferred from gain data in gaseous detectors through the so-called `feedback parameter' $\beta$. A quantitative description of this situation is indeed one of the missing pieces in the modelling and design of modern gaseous detectors. The $\beta$ parameter has been recently obtained in a number of Ar-mixtures through a full simulation including Penning transfer rates in \cite{Rob1, Rob2}. Under the assumption that $\beta$ is driven by photoelectric effect (either at the cathode or in the gas), it can be expressed as:
\beq
\beta\!\! =\!\!\! \int\!\!\!\!\int \!\! N_{ex}(E) \!\cdot\! \mathcal{P}_{scin} \!\cdot\! \frac{d^2P}{d\lambda d \Omega} (\lambda,\Omega) \!\cdot\! \mathcal{T}(\lambda,\Omega) \!\cdot\! QE(\lambda) d\lambda d\Omega \label{beta}
\eeq
where $\frac{dP}{d\lambda d \Omega}$ is the scintillation probability distribution, $\mathcal{T}$ is the medium transparency, $QE$ the photoelectric effect efficiency, and other magnitudes have been already defined. We have kept in eq. \ref{beta} the main dependencies and so $\mathcal{P}_{scin}$ is assumed to be field-independent, as observed in this work.
Inspection of eq. \ref{beta} shows that, upon variations of the concentration of quencher and if no re-emission is present, only i) the number of excited states, ii) the probability of scintillation and iii) the medium transparency can play a role in the maximum gain that can be achieved in a gaseous detector before photon feedback hampers operation. According to TDM, a $\beta \sim 1/\tau{K}f$ dependence would be expected if quenching would be the most important process, and this is precisely what has been observed for gaseous detectors filled with argon in combination with CH$_4$, C$_3$H$_8$ or C$_2$H$_2$ in \cite{Rob2}.

\section{Conclusions}

We have introduced a fully-microscopic electron transport model that can compute in detail both the time and wavelength spectra of the scintillation produced in xenon-based optical time projection chambers at typical operating conditions ($\lambda=[100$-$1500]$ nm). The model has been compared with time distributions measured for the xenon $2^{\tn{nd}}$ continuum in the pressure range 0.1-10bar, showing good agreement. The primary scintillation yields obtained in the absence of charge recombination correspond to $W_{sc}=40$eV (VUV) and $W_{sc}=86$eV (IR). Moreover, the simulation framework reproduces as well the measured scintillation in Xe-CO$_2$ and Xe-CF$_4$ mixtures, pointing to a dominant role of the xenon triplet-state in the scintillation process (TDM). This allows for instance to understand why CO$_2$ quenches argon scintillation 50 times more strongly than xenon one, a fact with obvious implications for the design of future optical TPCs.

Within the experimental systematics available, the scintillation model displays ambiguities related to plausible choices of the atomic and excimer pathways, that lead to high pressure extrapolations varying in the range $\times[1/2.5,1]$ for the concentrations of interest in future high pressure xenon experiments ($\sim 10$bar). Most remarkably, it was not possible to conclude unambiguously from the available data on the presence of termolecular reactions of the type Xe$^*$+Xe+M, for which existing information is very scarce and has not been replicated yet. A satisfactory description of the data could be achieved if such a mechanism would be simply ignored.

Additional support to the manifestation of the triplet dominance model (TDM) in earlier gaseous detectors' data when operated in Ar and Xe-based mixtures was provided.

\ack

DGD is supported by the Ramon y Cajal program (Spain). The authors want to acknowledge the RD51 collaboration for encouragement and support during the elaboration of this work, and in particular discussions with F. Resnati, A. Milov, V. Peskov, M. Suzuki and A. F. Borghesani.

The NEXT Collaboration acknowledges support from the following agencies and institutions: the European Research Council (ERC) under the Advanced Grant 339787-NEXT; the Ministerio de Econom\'ia y Competitividad of Spain under grants FIS2014-53371-C04 and the Severo Ochoa Program SEV-2014-0398; the GVA of Spain under grant PROM- ETEO/2016/120; the Portuguese FCT and FEDER through the program COMPETE, project PTDC/FIS-NUC/2525/2014 and UID/FIS/04559/2013; the U.S. Department of Energy under contracts number DE-AC02-07CH11359 (Fermi National Accelerator Laboratory) and DE-FG02-13ER42020
(Texas A\&M); and the University of Texas at Arlington.

\appendix
\section{Main rate constants for Xe/CO$_2$ mixtures}
The absolute 2-body ($K_{2}$) and 3-body $(K_{3}$) reaction rates in pure xenon are given in table \ref{TableI} at a pressure of 1bar (and $T=20^\circ$C), assuming a perfect gas. The sum of all transition coefficients from state $i$ to $j$ ($\sum_j A_{ij}$) is also given, together with the energy relative to the ground level of atomic xenon, and the Racah and Paschen notations for each state. Table \ref{TableII} shows the state-to-state population probabilities after 2-body reactions in pure xenon, and table \ref{TableIII} contains the parameters used for the VUV-emission pathway diagram plotted in Fig. \ref{TableIII}.

\onecolumn
\begin{table}[h]
\begin{minipage}{\textwidth}
  \centering
  \begin{tabular}{|c|c|c|c|c|c|}
     \hline
     ~ state (Paschen)\footnote{Paschen notation: the excited states are indexed in sequential order as $n' = n - N + l$, with $n, l$ being the principal and orbital quantum numbers of the valence electron, respectively, and $N$ is the principal quantum number of the electron in the ground state. Next, a letter $\lambda$ labels the orbital quantum number of the valence electron, with a sequential sub-index $i$ in inverse order of excitation energy. The shorthand expression for Paschen notation is $n'\lambda_i$. Exceptions to the general numbering scheme are indicated through $'$, $''$, etc.}
     ~ & ~ state (Racah)\footnote{Racah notation: the excited states are assimilated to a system formed by a core of electrons and a valence one (with well defined quantum numbers: $\vv{j} = \vv{l} + \vv{s}$, subindex $c$ and $v$ respectively). The shorthand expression for Racah notation is $n\lambda[K]_J$ where $K$ results from the composition $\vv{K} = \vv{j}_c + \vv{l}_v$, $J$ is the total angular momentum $\vv{J} = \vv{j}_c + \vv{j}_v$, $n$ is the principal quantum number of the valence electron and $\lambda$ the letter associated to its orbital momentum. Sub-multiplets that appear highly separated in energy may be indicated with $'$.}  ~ & ~energy [eV] ~& ~ $\sum_j A_{ij}$ [$\tn{ns}^{-1}$] ~ & ~ $K_{2}@1$bar [$\tn{ns}^{-1}$] ~ & ~ $K_{3}@1$bar [$\tn{ns}^{-1}$] \\
     \hline
     1s$_1$        &      -        & 0.000  & -                                   & -                        & -      \\
     1s$_5$        & $6s[3/2]_2$   & 8.315  & $2.33 \!\times\! 10^{-11}$            & $4.94\!\times\! 10^{-5}$ & 0.1465 \\
     1s$_4$        & $6s[3/2]_1$   & 8.437  & $0.281 / n_H$                         &           -              & 0.0855 \\
     1s$_3$        & $6s'[1/2]_0$  & 9.447  & $1.28\!\times\!10^{-8}$               & 0.2224                   &    -   \\
     1s$_2$        & $6s'[1/2]_1$  & 9.570  & $0.246 / n_H$                         & 2.4954                   &    -   \\
     2p$_{10}$     & $6p[1/2]_1$   & 9.580  & $0.026$                               & 3.7802                   &    -   \\
     2p$_9$        & $6p[5/2]_2$   & 9.686  & $0.027$                               & 2.7425                   &    -   \\
     2p$_8$        & $6p[5/2]_3$   & 9.721  & $0.031$                               & 1.8036                   &    -   \\
     2p$_7$        & $6p[3/2]_1$   & 9.789  & $0.028$                               & 4.3979                   &    -   \\
     2p$_6$        & $6p[3/2]_2$   & 9.821  & $0.036$                               & 2.0062                   &    -   \\
     3d$_6$        & $5d[1/2]_0$   & 9.890  & $4.36\!\times\!10^{-3}$               & 9.7649                   &    -   \\
     3d$_5$        & $5d[1/2]_1$   & 9.917  & $0.015 / n_H$                         & 4.8328                   &    -   \\
     2p$_5$        & $6p[1/2]_0$   & 9.933  & $0.031$                               & 0.1599                   & 0.4273 \\
     3d$_4'$       & $5d[7/2]_4$   & 9.943  & $4.34\!\times\!10^{-3}$               & 4.8676                   &    -   \\
     3d$_3$        & $5d[3/2]_2$   & 9.959  & $8.16\!\times\!10^{-3}$               & 4.8664                   &    -   \\
     3d$_4$        & $5d[7/2]_3$   & 10.039 & $7.34\!\times\!10^{-3}$               & 4.8510                   &    -   \\
     3d$_1''$      & $5d[5/2]_2$   & 10.157 & $1.21\!\times\!10^{-3}$               & 4.8649                   &    -   \\
     3d$_1'$       & $5d[5/2]_3$   & 10.220 & $1.39\!\times\!10^{-3}$               & 4.8639                   &    -   \\
     3d$_2$        & $5d[3/2]_1$   & 10.401 & $~3.04\!\times\!10^{-3} / n_H$        & 1.3637                   &    -   \\
     2s$_5$        & $7s[3/2]_2$   & 10.562 & $0.018$                               & 4.9415                   &    -   \\
     2s$_4$        & $7s[3/2]_1$   & 10.593 & $0.178 / n_H$                         & 4.9415                   &    -   \\
     3p$_{10-5}$\footnote{states from 3p$_{10}$ to 3p$_5$ are grouped, due to their proximity in energy.}
                   &      -        & 10.902	& $0.010$                               & 12.6008                  &    -   \\
     2p$_4$        & $6p[3/2]_1$   & 10.958 & $0.024$                               & 10.3277                  &    -   \\
     4d$_{10-6,4,3}$\footnote{states from 4d$_{10}$ to 4d$_6$ and 4d$_4$, 4d$_3$ are grouped, due to their proximity in energy.}
                   &      -    & 10.971 & $0.014$                               & 5.9298                   &    -   \\
     4d$_5$        & $6d[1/2]_1$   & 10.979 & $0.018$                               & 4.8426                   &    -   \\
     2p$_3$        & $6p[3/2]_2$   & 11.055 & $0.036$                               & 11.6125                  &    -   \\
     2p$_2$        & $6p[1/2]_1$   & 11.069 & $0.033$                               & 10.3277                  &    -   \\
     2p$_1$        & $6p[1/2]_0$   & 11.141 & $0.027$                               & 10.4018                  &    -   \\
     4d$_2$        & $6d[3/2]_1$   & 11.163 & $0.716 / n_H$                         & 4.8674                   &    -   \\
     Xe$^{**}$     &      -        & 11.7   & -                                     & 12.35                    &    -   \\
%     \hline
%     X$^1\Sigma_g$ &      -        &  -0.024  & - & - & - &\\
%     A$^3\Sigma_u$ &      -        &  7.92     & $0.01$ & 0.00013 & - &\\
%     A$^1\Sigma_u$ &      -        &  8.05     & $0.22$ & 0.0031 & - &\\
%     DSigma        &      -        &  9.53     & $1$ & - & - &\\
%     Xe$_2^{**}$   &      -        &  11.7     & 0.13 & $12.35$ & - &\\
%     $X0_g^+$      &      -        &  0.000    & - & - & - &\\
%     $A1_u/0_u^-$  &      -        &  8.315    & $0.025$ & 1.72 & - &\\
%     $A0_u^+$      &      -        &  8.437    & $0.2$ & 9.58 & - &\\
%     DSigma        &      -        &  9.74     & $1$ & - & - &\\
%     Xe$_2^{**}$(far bound)& -     &  11.7     & $0.20$ & 12.35 & - &\\
     \hline
   \end{tabular}
  \caption{Xenon atomic states in Paschen's and Racah's notation. The parameters shown are the sum of all radiative transition coefficients $\tau_i^{-1} = \sum_j A_{ij}$, and the binary $K_2$ and ternary $K_3$ reaction rates evaluated at 1bar ($T=20^o$C, perfect gas assumed). For resonant states the number of absorption-emission cycles is indicated as $n_H$. References are given in text.} \label{TableI}
\end{minipage}
\end{table}

\begin{table}[h]
  \centering
  \begin{tabular}{|c|c|c|c|c|c|c|c|c|c|c|}
     \hline
     -              & ~ 1s$_1$ ~ & ~ 1s$_5$ ~ & ~1s$_4$ ~& ~ 1s$_3$ ~ & ~ 1s$_2$ ~ & ~ 2p$_{10}$ ~ & ~ 2p$_9$ ~ & ~ 2p$_8$ ~ & ~2p$_7$ ~ & ~ 2p$_6$ ~\\
     \hline
     ~ 1s$_1$ ~     & ~ - ~      & ~ 0 ~      & ~ 0 ~    & ~ 0 ~      & ~ 0 ~      & ~ 0 ~         & ~ 0 ~      & ~ 0 ~      & ~ 0 ~     & ~ 0 ~\\
     ~ 1s$_5$ ~     & ~ 1$^{(1)}$ ~      & ~ - ~      & ~ 0 ~    & ~ 0 ~      & ~ 0 ~      & ~ 0 ~         & ~ 0 ~      & ~ 0 ~      & ~ 0 ~     & ~ 0 ~\\
     ~ 1s$_4$ ~     & ~ 0 ~      & ~ 0 ~      & ~ - ~    & ~ 0 ~      & ~ 0 ~      & ~ 0 ~         & ~ 0 ~      & ~ 0 ~      & ~ 0 ~     & ~ 0 ~\\
     ~ 1s$_3$ ~     & ~ 0 ~      & ~ $0.11^{(2,3)}$ ~   & ~ $0.89^{(2,3)}$ ~ & ~ - ~      & ~ 0 ~      & ~ 0 ~         & ~ 0 ~      & ~ 0 ~      & ~ 0 ~     & ~ 0 ~\\
     ~ 1s$_2$ ~     & ~ 0 ~      & ~ $0.010^{(2,3)}$ ~  & ~ $0.079^{(2,3)}$ ~& ~ $0.247^{(3)}$ ~  & ~ - ~      & ~ $0.663^{(4)}$ ~     & ~ 0 ~      & ~ 0 ~      & ~ 0 ~     & ~ 0 ~\\
     ~ 2p$_{10}$ ~  & ~ 0 ~      & ~ $0.014^{(3)}$ ~  & ~ $0.116^{(3)}$ ~& ~ $0.216^{(4)}$ ~  & ~ $0.654^{(4)}$ ~  & ~ - ~         & ~ 0 ~      & ~ 0 ~      & ~ 0 ~     & ~ 0 ~\\
     ~ 2p$_{9}$ ~   & ~ 0 ~      & ~ 0 ~      & ~ 0 ~    & ~ $0.3604^{(4)}$ ~ & ~ $0.1351^{(3)}$ ~ & ~ $0.405^{(4)}$ ~      & ~ - ~      & ~ $0.099^{(4)}$ ~  & ~ 0 ~     & ~ 0 ~\\
     ~ 2p$_{8}$ ~   & ~ 0 ~      & ~ 0 ~      & ~ 0 ~    & ~ $0.178^{(3)}$ ~  & ~ $0.110^{(3)}$ ~  & ~ $0.245^{(3)}$ ~     & ~ $0.466^{(3)}$ ~  & ~ - ~      & ~ 0 ~     & ~ 0 ~\\
     ~ 2p$_{7}$ ~   & ~ 0 ~      & ~ 0 ~      & ~ 0 ~    & ~ $0.348^{(3)}$ ~  & ~ $0^{(2)}$ ~      & ~ $0.011^{(2)}$ ~     & ~ $0.067^{(2)}$ ~  & ~ $0.539^{(2)}$ ~  & ~ - ~     & ~ $0.034^{(3)}$ ~\\
     ~ 2p$_{6}$ ~   & ~ 0 ~      & ~ 0 ~      & ~ 0 ~    & ~ $0.234^{(3)}$ ~  & ~ $0.001^{(2)}$ ~  & ~ $0.001^{(2)}$ ~      & ~ $0.345^{(3)}$ ~ & ~ $0.259^{(3)}$ ~  & ~ $0.161^{(3)}$ ~ & ~ - ~\\
     \hline
   \end{tabular}
  \caption{State-to-state quenching probabilities of the first 10 xenon excited states after Leichner\cite{Leichner}$^{(1)}$, Moutard\cite{Moutard}$^{(2)}$, Alford\cite{Alford92}$^{(3)}$ and Ku\cite{Setser86}$^{(4)}$. The total rate constants are given in Table \ref{TableI}.}\label{TableII}
\end{table}

\begin{table}[h]
\begin{minipage}{\textwidth}
  \centering
  \begin{tabular}{|c|c||c|c||c|c||c|c|}
     \hline
     ~ $\tau_{^3\!\Sigma}$\footnote{Hund's case $a$ notation: the axial projections of the angular momenta (along the internuclear axis) are good quantum numbers. The axial projection of the total angular momentum ($\Omega$) is used as the subindex of a capital letter indicating the axial projection of the orbital momentum ($\Lambda$). The spin ($s$), and waveform symmetries (reflection $P$ and inversion $I$) are included as additional super or sub-indexes). An additional prepended capital letter refers to the energy ordering of the state, in sequential order as $\mathds{N} = X, A, B, C, ...$. In the absence of rotational degrees of freedom, the shorthand expression for Hund's case $a$ notation is $ \mathds{N} ^{2s+1}\Lambda_{\Omega, I}^P$.}
     ~ & ~ 100 ns$^{(1)}$ ~ & ~ $K_{Q,^3\!\Sigma(M)}$ ~& ~ 11.12 ns$^{-1}$ ~ & ~ $K_{s_5\rightarrow1_u}$ ~ & ~ 0.1465 ns$^{-1}$$^{(1)}$     & ~ $K_{s_5\rightarrow1_u(M)}$ ~ & ~ 116 ns$^{-1}$ ~\\
%     \hline
     ~ $\tau_{^1\!\Sigma}$ ~ & ~ 4.55 ns$^{(1)}$  ~ & ~ $K_{Q,^1\!\Sigma(M)}$ ~& ~ 12.85 ns$^{-1}$ ~ & ~ $K_{s_4\rightarrow0_u}$ ~ & ~ 0.0855 ns$^{-1}$$^{(1)}$     & ~ $K_{s_4\rightarrow0_u(M)}$ ~ & ~ 116 ns$^{-1}$$^{(6)}$ ~\\
     ~ $\tau_{1_u}$\footnote{Hund's case $c$ notation: in the absence of rotational degrees of freedom only the axial projection of the total angular momentum (along the internuclear axis) is a good quantum number $\Omega$. The waveform symmetries (reflection $P$ and inversion $I$) are included as additional super and sub-indexes, respectively). An additional prepended capital letter refers to the energy ordering of the state, in sequential order as $\mathds{N} = X, A, B, C, ...$. In the absence of rotational degrees of freedom, the shorthand expression for Hund's case $c$ notation is $ \mathds{N} \Omega_{I}^P$.}
     ~ & ~ 40 ns$^{(1)}$  ~ & ~ $K_{Q,1_u(M)}$ ~& ~ 11.12 ns$^{-1}$ ~ & ~ $K_{0_u\rightarrow{s_4}}$ ~ & ~ 1.43 ns$^{-1}$$^{(1)}$     & ~  & ~ \\
     ~ $\tau_{0_u}$        ~ & ~ 5 ns$^{(1)}$  ~ & ~ $K_{Q,0_u(M)}$ ~& ~ 12.85 ns$^{-1}$ ~ & ~ $K_{0_u\rightarrow{s_5}}$ ~ & ~ 6.42 ns$^{-1}$$^{(1)}$     & ~  & ~ \\
     ~ $\tau_{s_5}$        ~ & ~ 42 s$^{(2)}$  ~ & ~ $K_{Q,s_5(M)}$ ~& ~ 11.12 ns$^{-1}$$^{(4)}$ ~ & ~ $K_{cool,0_u}$ ~ & ~ 1.72 ns$^{-1}$$^{(1)}$ & ~ & ~ \\
     ~ $\tau_{s_4}$        ~ & ~ $3.56\times n_H$ ns$^{(3)}$  ~ & ~ $K_{Q,s_4(M)}$ ~& ~ 12.85 ns$^{-1}$$^{(5)}$ ~ & ~ $K_{cool,1_u}$ ~ & ~ 1.72 ns$^{-1}$$^{(1)}$ & ~ & ~ \\
\hline
\end{tabular}
  \caption{Pathways determining the emission of the $1^{\tn{st}}$ and $2^{\tn{nd}}$ continuum in Xe/CO$_2$ mixtures after Moutard\cite{Moutard}$^{(1)}$, Walhout\cite{Walhout}$^{(2)}$, NIST\cite{NIST}$^{(3)}$, Velazco\cite{Setser78}$^{(4)}$, Alekseev\cite{Alekseev96}$^{(5)}$ and Wojciechowski\cite{Wojche}$^{(6)}$. Rate constants ($K$) are evaluated at 1 bar (and $T=20^\circ$C) by assuming a perfect gas. For the molecular states the 2-body quenching rates are not known (3$^{\tn{rd}}$-4$^{\tn{th}}$ columns) so the ones of the s$_4$ or s$_5$ states are used instead. The 3-body quenching rates for the 1s$_4$ and 1s$_5$ states in the presence of CO$_2$ are assumed to be identical (last two columns) and it is further assumed that this 3-body channel helps at stabilizing the Xe$_2^*$ excimer formation, as argued by authors in \cite{Firestone} and \cite{Wojche}. For details on this assumption see text. A schematic diagram is shown in Fig. \ref{Decay2ndScheme}. Sub-indexes labeling states refer to Hund's case $a$ and $b$ notations.}\label{TableIII}
\end{minipage}
\end{table}

\twocolumn

\section{Analytical approximation for VUV emission in xenon mixtures} \label{analyticalPathways}

The decay scheme shown in Fig. \ref{Decay2ndScheme} admits a direct analytical solution, that has been given in text, as:

\bear
&&\mathcal{P}_{scin, s_4}\!=\!\frac{\mathcal{P}_{pop,s_4} \cdot \mathcal{P}_{s_4\rightarrow{0_u}} }{1-\mathcal{P}_{s_4\rightarrow 0_u} \cdot \mathcal{P}_{0_u\rightarrow s_4 }} \times \label{ana3}  \nonumber \\
&& ~\left( \mathcal{P}_{cool,0_u} \!\!\cdot\! \mathcal{P}_{rad,^1\Sigma} \!+\! \mathcal{P}_{0_u\rightarrow{s_5}} \!\!\cdot\! \mathcal{P}_{s_5\rightarrow{1_u}} \!\!\cdot\! \mathcal{P}_{cool,1_u} \!\!\cdot\! \mathcal{P}_{rad,^3\Sigma} \!\right) \\
&&\mathcal{P}_{scin, s_5}\!=\!\mathcal{P}_{pop, s_5} \!\cdot\! \mathcal{P}_{s_5\rightarrow{1_u}} \!\cdot\! \mathcal{P}_{cool,1_u} \!\cdot\! \mathcal{P}_{rad,^3\Sigma} \label{ana4}
\eear

The corresponding terms can be expressed as follows. The population probabilities must be obtained from the full simulation, however a good approximation (that is followed in text) is:
\bear
\mathcal{P}_{pop,s_4} \simeq 0.5 \\
\mathcal{P}_{pop,s_5} \simeq 0.5
\eear
The probabilities of excimer formation via 3-body reactions can be expressed as:
\bear
&&\mathcal{P}_{s_4\rightarrow{0_u}}= \nonumber\\
&&\frac{(1\!-\!f)^2n^2K_{s_4\rightarrow{0_u}} \!+ rfn^2K_{Q,s_4(M)}}{(1\!-\!f)^2n^2K_{s_4\rightarrow{0_u}} \! +\! rfn^2K_{s_4\rightarrow{0_u}(M)} \!+ \!fnK_{Q,s_4(M)}} \\
&&\mathcal{P}_{s_5\rightarrow{1_u}}= \nonumber\\
&&\frac{(1\!-\!f)^2n^2K_{s_5\rightarrow{1_u}} \!+ rfn^2K_{Q,s_5(M)}}{(1\!-\!f)^2n^2K_{s_5\rightarrow{1_u}} \! + \! rfn^2K_{s_5\rightarrow{1_u}(M)} + fnK_{Q,s_5(M)}}
\eear
where f is the additive concentration, $n=P/P_o$ the pressure relative to $P_o=1$ bar and $r$ indicates the fraction of
3-body reactions with the additive that contributes to excimer formation. Throughout the text it has been assumed $r=1$.
The rate constants are evaluated at 1bar, the temperature is assumed constant and the mixture is assumed to behave like a perfect gas.

The cooling probabilities can be expressed as:
\bear
&&\mathcal{P}_{cool,0_u} = \frac{(1\!-\!f)nK_{cool,0_u}}{K_{0_u, total}}\\
&&\mathcal{P}_{cool,1_u} = \frac{(1\!-\!f)nK_{cool,1_u}}{K_{1_u, total}}\\
\eear
with:
\bear
K_{0_u, total} = (1\!-\!f)nK_{cool,0_u} + (1\!-\!f)nK_{0_u\rightarrow{s_5}} +...\nonumber\\
(1\!-\!f)nK_{0_u\rightarrow{s_4}} + fnK_{Q,0_u(M)} + 1/\tau_{0_u} \\
K_{1_u, total} = (1\!-\!f)nK_{cool,1_u} + fnK_{Q,1_u(M)} + 1/\tau_{1_u}
\eear

The dissociative probabilities of the $0_u$ state are:
\bear
&&\mathcal{P}_{0_u\rightarrow{s_4}} = \frac{(1\!-\!f)nK_{0_u\rightarrow{s_4}}}{K_{0_u, total}} \\
&&\mathcal{P}_{0_u\rightarrow{s_5}} = \frac{(1\!-\!f)nK_{0_u\rightarrow{s_5}}}{K_{0_u, total}}
\eear
and finally the radiative probabilities for the singlet and triplet state are standard quenching relations:
\bear
\mathcal{P}_{rad,^1\Sigma} = \frac{1/\tau_{_{^1\Sigma}}}{1/\tau_{_{^1\Sigma}} + fnK_{Q,^1\Sigma}} \\
\mathcal{P}_{rad,^3\Sigma} = \frac{1/\tau_{_{^3\Sigma}}}{1/\tau_{_{^3\Sigma}} + fnK_{Q,^3\Sigma}}
\eear

Evaluation of these formulas for the values tabulated in \ref{TableIII} can be found in Fig. \ref{AnalyticPathways}.

\section{Relation between attachment coefficient and light fluctuations} \label{attachSection}

In the limit of small attachment ($\eta$) the distribution function of the light yield ($Y$) for an individual electron crossing a gas gap ($g$) can be approximated by:\footnote{Throughout the text we have referred to $Y$ as the `optical gain' $Y\equiv m_\gamma$.}
\beq
\frac{dN}{dY} = \eta g \frac{[\Theta(Y) - \Theta(Y-Y_o)]}{Y_o} + (1-\eta g) \delta(Y-Y_o) \label{Ana1}
\eeq
The first term represents the probability that the electron is attached during its transit through the gap, and the second
one the probability that it reaches the anode. The latter has a certain natural width but, as discussed in text,
it is too small to become important and can be approximated by a $\delta$ function. The yield in absence of attachment is expressed as $Y_o$ and $\Theta$ is the step function. Realizing that the scintillation probability distribution in the presence of a mild attachment can be expressed as Eq. \ref{Ana1} is all that is needed to derive two useful relations.
The mean value can be written as:
\bear
&& \bar{Y} = \int_o^\infty \frac{dN}{dY} Y dY = \eta g \int_o^{Y_o} \frac{Y}{Y_o} dY + (1-\eta g) Y_o \label{eq1} \\
&& \bar{Y} = (1-\frac{1}{2}\eta g) Y_o
\eear
The variance:
\bear
\sigma_Y^2 = \!\!\! && \int_o^\infty \!\frac{dN}{dY} (Y\!-\!\bar{Y})^2 dY = \eta g \!\! \int_o^{Y_o} \!\frac{(Y-\bar{Y})^2}{Y_o} dY \!+\! ... \\
&& (1-\eta g) (Y_o - \bar{Y})^2 \label{eq1}
\eear
After some algebra, the $Q$-factor can be written as:
\bear
Q = && \frac{\sigma_Y^2}{\bar{Y}^2} = \frac{1}{(1-\frac{1}{2}\eta g)} \bigg[ \frac{1}{24}(\eta g)^4 + ... \\
    && \frac{1}{3}(\eta g) (1-\frac{1}{2}\eta g)^3 + \frac{1}{4}(\eta g)^2 - \frac{1}{4}(\eta g)^3 \bigg] = \\
    && \frac{1}{3}\eta g + O((\eta g)^2)
\eear
that is the result used in text.

\end{document}